\documentclass[aps,prb,twocolumn,superscriptaddress,amsfonts,citeautoscript,nofootinbib]{revtex4-2}
\usepackage{amsmath, amssymb, graphicx, color, xcolor, braket,bbold,multirow,mathtools,verbatim,float}
\usepackage[urlcolor=blue,colorlinks=true,citecolor=blue,linkcolor=blue,pdfstartview={FitH},bookmarks=false]{hyperref}

\makeatletter
\def\bib@control@title{yes}
\makeatother

\begin{document}

\title{{Andreev bound state spectroscopy of a quantum-dot-based  Aharonov-Bohm interferometer with superconducting terminals}}

\author{Peter Zalom}
\email{zalomp@fzu.cz}
\affiliation{Institute of Physics, Czech Academy of Sciences, Na Slovance 2, CZ-18200 Praha 8, Czech Republic}

\author{Don Rolih}
\email{don.rolih@ijs.si}
\affiliation{Jo\v{z}ef Stefan Institute, Jamova 39, SI-1000 Ljubljana, Slovenia}
\affiliation{Faculty of Mathematics and Physics, University of Ljubljana, Jadranska 19, SI-1000 Ljubljana, Slovenia}

\author{Rok \v{Z}it\textbf{}ko}
\email{rok.zitko@ijs.si}
\affiliation{Jo\v{z}ef Stefan Institute, Jamova 39, SI-1000 Ljubljana, Slovenia}
\affiliation{Faculty of Mathematics and Physics, University of Ljubljana, Jadranska 19, SI-1000 Ljubljana, Slovenia}

\date{\today}

\begin{abstract}
We analytically and numerically investigate an Aharonov-Bohm interferometer with two superconducting terminals and a strongly correlated quantum dot in one arm. Through a rigorous derivation, we prove that this double-path interferometer is spectrally equivalent to a simpler system: an interacting quantum dot coupled to a non-interacting side-coupled proximitized mode and a semiconductor lead. This equivalence reveals a simple interpretation of the interferometer's behavior through the competition of a geometric factor $\chi$, a key parameter characterizing the anomalous part of the hybridization function, with the properties of the side-coupled mode. We identify the conditions for the formation of doublet chimney in the phase diagrams in more general setting. Moreover, we show how the obtained Andreev bound state spectra clearly indicate the presence of Josephson diode effect generated by interferometric phenomena.
\end{abstract}


\maketitle

\section{Introduction \label{sec:intro}}

The Aharonov-Bohm (AB) effect--whereby charged particles traversing different paths enclosing a magnetic flux acquire a relative phase shift--fundamentally connects quantum mechanics with electromagnetism. First predicted by Aharonov and Bohm in 1959 \cite{Aharonov-1959}, it was almost immediately verified by Chambers using electron beams in an electron microscope \cite{Chambers-1960}. In mesoscopic and nanoscopic physics, the first observations were made via conductance oscillations in metallic rings \cite{Webb-1985,Chandrasekhar-1991}. Early theoretical works employed scattering theory \cite{Stone-1986,Washburn-1986,Beenakker-1997,vanOudenaarden-1997,vanOudenaarden-1998} or Green's function techniques \cite{Konig-2002}. Later, the numerical renormalization group (NRG) approaches expanded the focus towards more complex situations such as quantum dots (QDs) embedded in one or both paths of AB rings \cite{Pascal-2005-ab,Kubala-2014}. 

AB ring interferometers with superconducting (SC) leads received relatively little attention. Research primarily focused on hybrid metal-superconductor cases \cite{Blasi-2022} and purely SC scenarios \cite{Nazarov-1993-absc,Nazarov-2009,Karrasch-2009-ab,Yi-2010,Dolgirev-2019} with significant interest in Fano effects \cite{Zhang-2005,Osawa-2008,Valkov-2019}. A new research direction was recently introduced in Ref.~\cite{Souto-2022}, where nonsinusoidal current-phase relations (CPRs) in interferometers composed of Josephson junctions (with or without interacting QDs) were explored for the design of Josephson diodes. The interplay between superconductivity and strong electron correlations, however, gives rise to some of the most challenging quantum many-body problems in mesoscopics and nanophysics. 

\begin{figure}[H]
	\includegraphics[width=1.0\columnwidth]{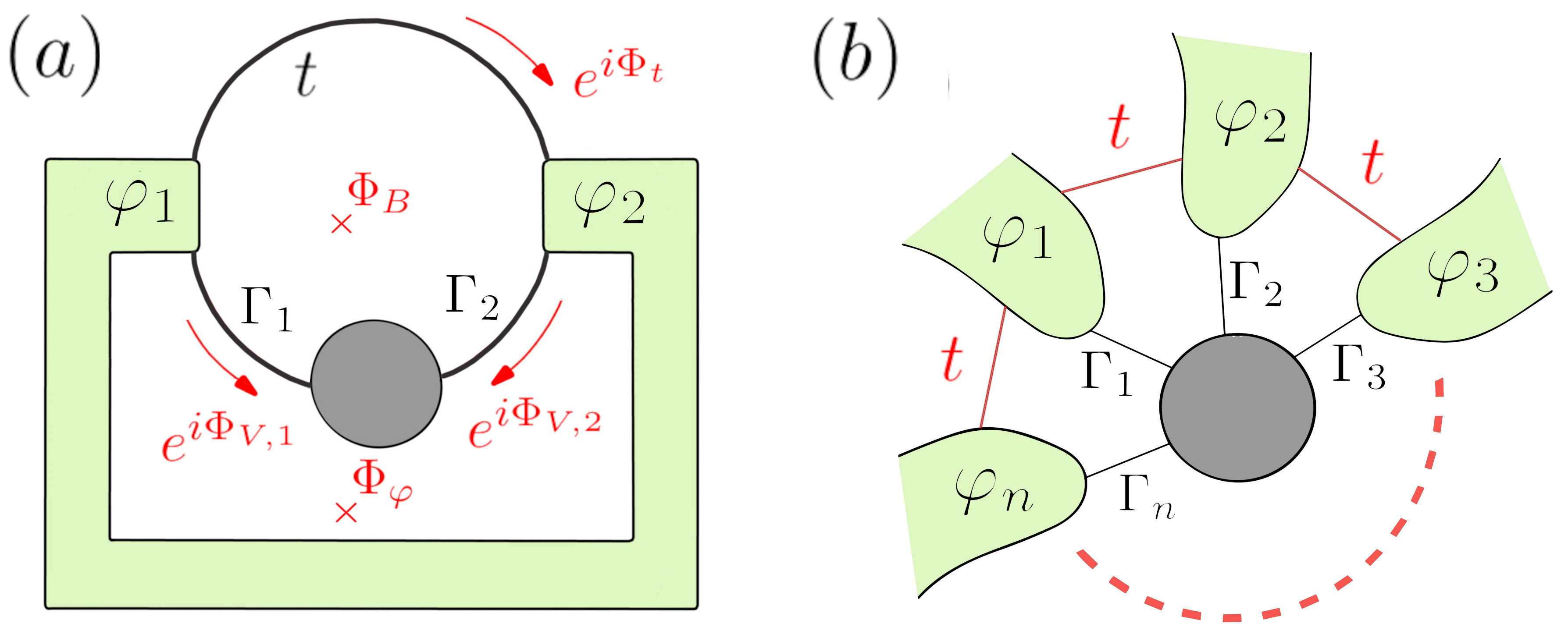}
	\caption{
		$(a)$
		Two-terminal interferometric device forming an Aharonov-Bohm ring with a direct electron hopping $t$ with phase $\Phi_t$ in one arm and a quantum dot (gray) in the other arm. The phases of electron hopping onto the dot are $\Phi_{V,1}$ and $\Phi_{V,2}$. Red arrows show the hopping phase orientations. Magnetic flux $\Phi_B$ pierces the ring, while $\Phi_{\varphi}$ pierces the superconducting loop (green).  
		$(b)$
		General $n$-terminal superconducting Anderson impurity model with the direct hopping $t$ (red) between the superconducting leads (green) and a centrally-placed strongly-interacting dot (gray).
		\label{fig:ab_ring}}
\end{figure}

Even for the simplest case of a QD embedded between two SC leads, an exact solution requires advanced numerical tools such as the NRG \cite{Satori-1992,Yoshioka-2000} or quantum Monte Carlo (QMC) \cite{Siano-2004,Luitz-2010,Luitz-2012}. Simplified approaches rely on effective models such as the atomic limit (AL) \cite{Meng-2009} or its generalization, GAL \cite{Zonda-2023}, and the zero-band-width approximation \cite{Grove-Rasmussen-2018}. More advanced expansions in terms of additional modes are available within the surrogate model \cite{Baran-2023} or the approach of Ref.~\cite{Bobok-2025}. The semi-analytic method of functional renormalization group (FRG) \cite{Kopietz-2010,Meden-2019,Karrasch-2008} has also been employed. These studies reveal the presence of singlet-doublet quantum phase transitions (QPTs) accompanied by Josephson current discontinuities and crossings of $\varphi$-dependent in-gap bound states, known as Yu-Shiba-Rusinov (YSR) or Andreev bound states (ABSs).

Focusing now on the AB effect in the presence of superconductivity and strong electronic correlations, we consider an interferometric device sketched in Fig.~\ref{fig:ab_ring}$(a)$. Single charge transport occurs both via direct tunneling and through the QD. The two paths enclose the magnetic flux $\Phi_B$, while the flux $\Phi_{\varphi}$ threading the SC loop primarily controls the BCS phase bias across the system. The two-terminal interferometric scenario is actually a special case of the general $n$-terminal variant, see Fig.~\ref{fig:ab_ring}$(b)$, which is known to host topologically distinct phases for $n \geq 3$ \cite{Klees-2020}. Even in the two-terminal case, a rich interplay of energy scales emerges, involving the SC gap $\Delta$, Coulomb repulsion $U$, level energy $\varepsilon_d$, direct hopping amplitude $t$, the coupling strengths $\Gamma_1$, $\Gamma_2$, and the various phases stemming from both the SC order parameters and hopping amplitudes. It is important to stress that similar interferometric devices have already experimentally demonstrated the appearance of controlled 0-$\pi$ transitions \cite{Kim-2011,Maurand-2012,Szombati-2016}. 

To examine such a complex system, Wilsonian renormalization group (RG) offers an excellent framework. However, so far only its semi-analytic variant FRG has been applied to this problem. While FRG revealed a complex landscape of singlet-doublet phases \cite{Karrasch-2009-ab}, it did not address all phases present in the system. Moreover, the truncation inherent in FRG, required to manage the infinite hierarchy of flow equations, breaks important system symmetries. In contrast, the NRG offers an exact numerical realization of the same Wilsonian RG approach. Its original implementation for gapped systems by Satori et al.~\cite{Satori-1992} was tailored specifically for BCS superconductors. However, in case of multiple superconducting leads with direct hopping between them, and in the presence of non-trivial phases, the lack of symmetries easily pushes the computational costs to the feasibility threshold, if not beyond.

In this work, we employ the novel log-gap NRG scheme~\cite{Liu-2016,Zalom-2023} which overcomes these limitations by formulating the problem in the framework of Anderson impurity orbital immersed in a bath with a gapped hybridization. The key innovation lies in devising a single semi-infinite chain independent of the number or nature of terminals. The chain is optimized to best represent the hybridization self-energy. Coupling this chain to the interacting Anderson impurity orbital allows an iterative diagonalization using a specific double-scaling law, so that a finite number of eigenstates can be retained throughout the calculation without compromising the RG flow of many-body spectra~\cite{Zalom-2023}.

While our ultimate goal is to obtain the numerically exact behavior of the ABSs, we first conduct a thorough analytic study. Our approach leverages the properties of the hybridization self-energy, which contains branch-cut and pole contributions: the former yield the usual hybridization function (defined as the jump discontinuity), while the latter give rise to non-interacting modes that are side-coupled to the interacting quantum dot. As a key simplification step, the analysis of the continuous part makes use of the recently introduced concepts of the geometric factor $\chi$ \cite{Zalom-2024} and the Bogolyubov-Valatin transformations of the quantum dot Hamiltonian \cite{Zalom-2021}. 

The resulting hybridization function resembles that of a semiconductor with varying particle-hole asymmetry governing the possible emergence of QPTs, revealing thus an analogy with the two-terminal superconducting Anderson impurity nodel (SC-AIM). Our analytic findings demonstrate that in interferometric SC two-terminal devices the in-gap states arise from the competition between the side-coupled mode and the symmetry encoded in the semiconductor hybridization function. Specifically, the symmetric hybridization function in combination with half-filled quantum dot leads to the emergence of the "doublet chimney"  in the phase diagram~\cite{Pavesic-2024}. We show that such special points emerge also in the AB ring structure and are of significance for qualitative understanding of the observed QPTs.



We will first review the concept of the geometric factor $\bm{\chi}$ describing the tunneling density of states and the role of Bogolyubov-Valatin transformation for the interpretation of the $\varphi$-dependent behavior of the two-terminal SC-AIM in Sec.~\ref{subsec:two-term} [see also Fig.~\ref{fig:sciam}]. Subsequently, we explore the parameter space of the two-terminal interferometer in Sec.~\ref{subsec:interf}.
The core of our analytic study is presented in Sec.~\ref{sec:analytic}. Five main steps can be identified, as visualized in Fig.~\ref{fig:overview}. A separate subsection in Sec. ~\ref{sec:analytic} is devoted to each step: the subsection \ref{subsec:eom}  describes the application of the equation of motion (EOM) technique, \ref{subsec:self_energy_2} defines the continuous and pole contributions appearing in the hybridization self-energy, and \ref{subsec:chi} discusses the BCS gauge invariance within the context of the geometric factor $\bm{\chi}$. Furthermore, the pole contribution of the self-energy is shown to give rise to a side-coupled mode attached to the interacting QD in~\ref{subsec:side_mode}, and the role of Bogolyubov-Valatin transformations is discussed in \ref{subsec:bogoliubov_valatin}. The consequences of our analytic findings are explored in Sec.~\ref{sec:extended} using a simple two-dot approximation that takes into account solely the effect of the side-coupled dot. Finally, the ABS spectroscopy is performed exactly by employing the log-gap NRG in Sec.~\ref{sec:nrg}.

%
%
%
%
%


\section{Theory \label{sec:chimney}}

\subsection{Two-terminal SC-AIM \label{subsec:two-term}}

\begin{figure}[t]
	\includegraphics[width=0.8\columnwidth]{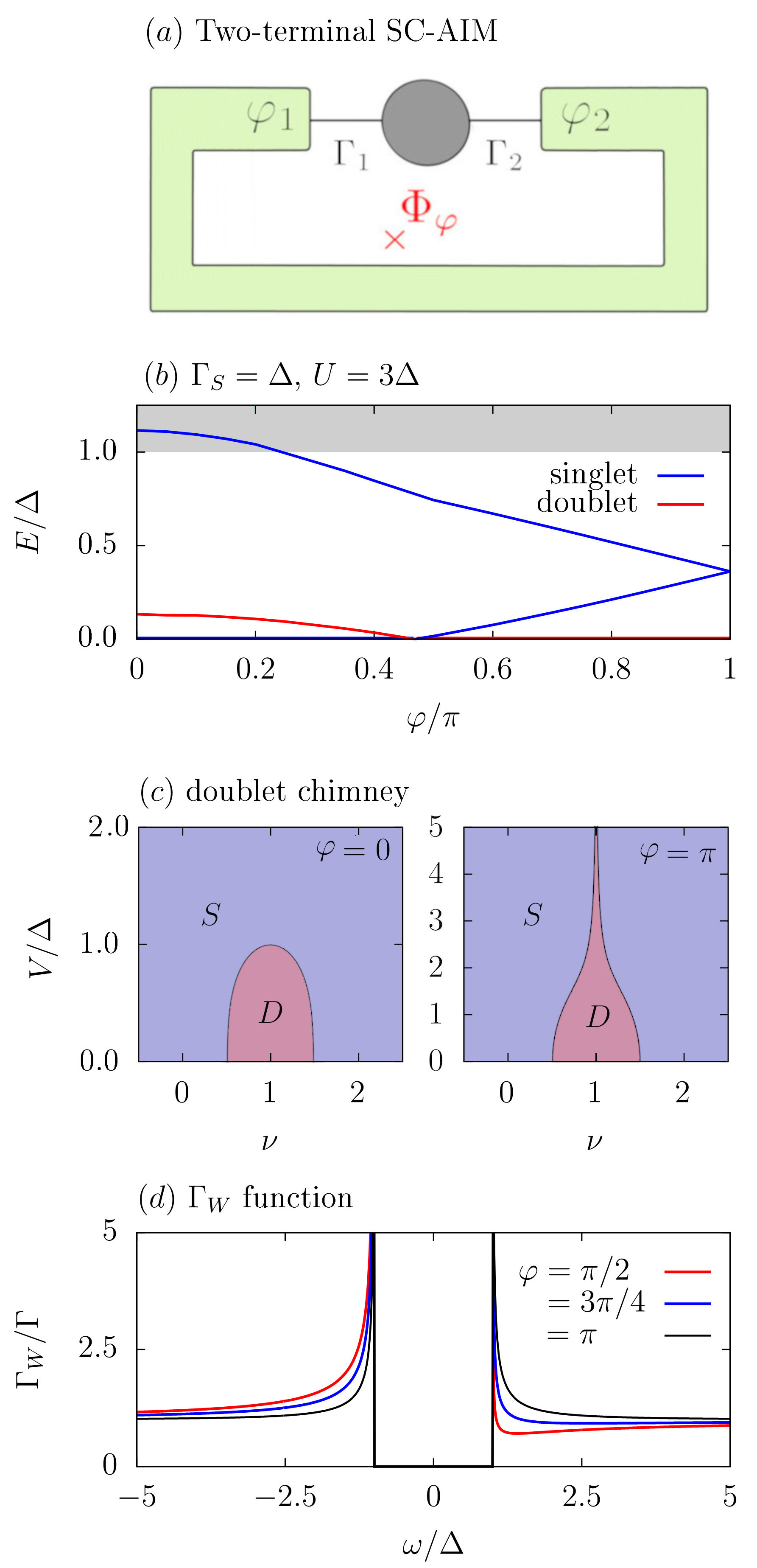}
	\caption{
		$(a)$
		Schematics of a two-terminal SC-AIM with BCS leads (green) and a strongly-interacting single level QD (gray). Magnetic flux $\Phi_{\varphi}$ controls the phase difference $\varphi \equiv  \varphi_2-\varphi_1 + \phi_{V,2} - \phi_{V,1}$.
		$(b)$
        Phase-dependence of the energy $E$ of the sub-gap states for $U/\Delta=3$, $\Gamma/\Delta=1$, $\varepsilon_d=-U/2$.
        $(c)$
		At $\varphi=\pi$, singlet (letter S) and doublet (D) ground states form a distinct pattern of a doublet chimney when plotted as $\nu \equiv 1/2-\varepsilon_d/U$ vs $V \equiv V_1 = V_2$ (data according to the minimal model of \cite{Pavesic-2024} with $U/\Delta = 10$ and $t/\Delta = 0.2$).
        $(d)$
        Hybridization functions $D^W(\omega)$ at $\varphi/\pi=0, 0.75, 1.00$ in the insulator basis of Ref.~\cite{Zalom-2021}.
		\label{fig:sciam}}
\end{figure}

To establish a baseline for the later analysis of the interferometric system, we first review its simpler single-pathway counterpart with only one arm [Fig.~\ref{fig:sciam}$(a)$], i.e., the standard two-terminal SC-AIM model:
\begin{align}
	H
	=
	H_d
	+
	\sum_j \left( H_{j,SC} + H_{j,T} \right)
	\label{eq:SC-AIM}
\end{align}
with $j \in \{ 1,2\}$, where
\begin{align}
	H_d
	&=
	\sum_{\sigma} 
	\varepsilon_{d}
	d^{\dagger}_{\sigma}
	d^{\vphantom{\dagger}}_{ \sigma}
	+
	U
	d^{\dagger}_{\uparrow}
	d^{\vphantom{\dagger}}_{ \uparrow}
	d^{\dagger}_{\downarrow}
	d^{\vphantom{\dagger}}_{ \downarrow},
	\label{eq:dotH}
	\\
	H_{j,SC}
	&=
	\sum_{\mathbf{k} j\sigma} 
	\, \varepsilon_{\mathbf{k}}
	c^{\dagger}_{\mathbf{k} j\sigma}
	c^{\vphantom{\dagger}}_{\mathbf{k} j\sigma}
	+
	\left(
	\Delta_{j}
	c^{\dagger}_{\mathbf{k}j \uparrow} 
	c^{\dagger}_{-\mathbf{k}j \downarrow}
	+
	\textit{H.c.}\right),
	\label{eq:bcsH}
	\\
	H_{j,T} 
	&= 
    \frac{1}{\sqrt{N}}
	\sum_{\mathbf{k} \sigma} \,
	\left(
	V_{\mathbf{k}j}^* 
	d^{\dagger}_{\sigma}
	c^{\vphantom{\dagger}}_{\mathbf{k}j\sigma}
	+
	V_{\mathbf{k}j}  
	c^{\dagger}_{\mathbf{k}j\sigma}
	d^{\vphantom{\dagger}}_{\sigma} 
	\right).
	\label{eq:tunnelH}
\end{align}
Here $c^{\dagger}_{\mathbf{k} j\sigma}$ ($c^{\vphantom{\dagger}}_{\mathbf{k}j\sigma}$) creates (annihilates) an electron of spin $\sigma \in \{\uparrow, \downarrow \}$ and quasi-momentum $\mathbf{k}$ in lead $j$, while $d^{\dagger}_{\sigma}$ ($d^{\vphantom{\dagger}}_{\sigma}$) performs the same operations for electrons on the QD. 
We assume identical SC materials for both leads, set therefore $\varepsilon_{\mathbf{k}j} \equiv \varepsilon_{\mathbf{k}}$ and equal gap parameters $\Delta \equiv |\Delta_j|$. Consequently, only the BCS phases vary so that $\Delta_{j} \equiv \Delta e^{i\varphi_{j}}$. The tunneling terms involve amplitudes $V_{\mathbf{k}j}$, which we for simplicity treat as $\mathbf{k}$-independent but complex-valued: $V_{\mathbf{k}j} \equiv V_{j} = |V_j| e^{-i \phi_{V,j}/2}$. $N$ is the number of levels in each contact. Although Hamiltonian \eqref{eq:SC-AIM} contains four phase variables explicitly, gauge invariance allows unitary transformations $c^{\dagger}_{\mathbf{k}j\sigma} \rightarrow c^{\dagger}_{\mathbf{k}j\sigma} e^{i \phi_{V,j}/2}$. These turn the hopping amplitudes $V_{j}$ to real-valued ones, while gap parameters absorb the hopping phases $\phi_{V,j}$ as $\Delta e^{i\varphi_{j}} \rightarrow \Delta e^{i(\varphi_{j} + \phi_{V,j})}$. Consequently, SC-AIM depends physically on a single gauge-invariant phase difference
\begin{align}
\varphi 
\equiv 
\varphi_2-\varphi_1 + \phi_{V,2} - \phi_{V,1}.
\end{align}

Because SC-AIM is not analytically solvable in an exact way, numeric approaches like NRG are applied to obtain the position of bound states emerging in the gap of SC-AIM. Their crossings at zero energy indicate QPTs as shown in Fig.~\ref{fig:sciam}(b) for $U/\Delta=3$ and $U/\Gamma=1$ at half filling, $\varepsilon_d=-U/2$. Most importantly, at $\varphi=\pi$, the half-filling condition ensures an underlying symmetry \footnote{The system is symmetric under $d_{\uparrow} \to d_{\downarrow}$, $d_{\downarrow} \to d_{\uparrow}$ followed by complex conjugation. This operation reflects spin across the $(xy)$ plane in the spin space. It is an anti-unitary symmetry which differs from the time-reversal symmetry inverting the spins at $\varphi = 0$, where $d_\uparrow \to -d_\downarrow$, $d_\downarrow \to d_\uparrow$.} which guarantees doublet GS even when all remaining parameters of SC-AIM are varied -- giving rise to the ``doublet chimney'' (persistent doublet state even in the strong-coupling limit) as shown within the minimial model of Ref.~\cite{Pavesic-2024} in Fig.~\ref{fig:sciam}(c), right panel. 

This behavior is rooted in the properties of the environment as can be revealed using the formalism presented in Refs.~\cite{Zalom-2021,Zalom-2022,Zalom-2024}. One introduces the Nambu formalism via spinors $D^{\dagger} =  \left(d^{\dagger}_{\uparrow}, 
d^{\vphantom{\dagger}}_{\downarrow} \right)$ 
and
$C^{\dagger}_{\alpha \, \mathbf{k}} 
=
\left(
c^{\dagger}_{\alpha \, \mathbf{k} \uparrow},
c^{\vphantom{\dagger}}_{\alpha \, -\mathbf{k} \downarrow}
\right)$. Employing equation of motion (EOM) techniques one integrates out the environmental degrees of freedom and obtains the hybridization self-energy that enters the local Green's function of the dot. For an infinitely wide band it reads \cite{Zalom-2024}:
\begin{align}
\mathbb{\Sigma}(z)
=
\Gamma
\frac{\sqrt{\Delta^2-z^2}}{z^2-\Delta^2}
\left(
\begin{array}{cc}
	z & -\Delta \bm{\chi} \\
	-\Delta \bm{\chi}^* & z
\end{array},
\right),
\end{align}
where $\bm{\chi}$ is a gauge-dependent complex-valued quantity given as $\bm{\chi} = \Gamma_1 e^{i(\varphi_1 + \phi_{V,1})}/\Gamma  + \Gamma_2 e^{i(\varphi_2 + \phi_{V,2})}/\Gamma$ with $\Gamma=\Gamma_1+\Gamma_2$. By performing the gauge transformation  
$
\varphi_i
\rightarrow 
\varphi_i - \arg(\bm{\chi})
$
for $i \in \{1,2\}$, $\bm{\chi}$ is rotated into the direction of the real axis, i.e., $\bm{\chi} \rightarrow \chi \equiv |\bm{\chi}|$~\cite{Zalom-2024}. Consequently:
\begin{align}
\mathbb{\Sigma}(z)
\rightarrow
\mathbb{\Sigma}(z)
=
\Gamma
\frac{\sqrt{\Delta^2-z^2}}{z^2-\Delta^2}
\left(
\begin{array}{cc}
	z & -\Delta \chi \\
	-\Delta \chi & z
\end{array}
\right),
\end{align}
which has exactly the same form as the coupling-symmetric scenario with $\Gamma/2 \equiv \Gamma_1=\Gamma_2$ \cite{Kadlecova-2017,Zalom-2024}. The corresponding hybridization function $\mathbb{D}(\omega)$ is then obtained from the jump (discontinuity) of the self-energy across the real axis as
\begin{align}
\mathbb{D}(\omega)
=
-
\frac{\mathbb{\Sigma}(\omega^+) - \mathbb{\Sigma}(\omega^-)}{2i}
\label{eq:hybridization_func}
\end{align}
with $\omega^+\equiv\omega+i\eta$ and $\omega^-\equiv\omega-i\eta$, $\omega$ being purely real and $\eta$ infinitesimally small. Using
\begin{align}
	-
	\frac{\sqrt{\Delta^2-(\omega^+)^2} - \sqrt{\Delta^2-(\omega^-)^2}}{2i} 
	=
	\text{sgn}(\omega) \sqrt{\omega^2-\Delta^2},
    \label{eq6}
\end{align}
we obtain a matrix-valued purely real-valued hybridization function
 \begin{align}
	\mathbb{D}(\omega)
	&=
	\frac{\Gamma \text{sgn}(\omega) }
	{ \sqrt{\omega ^2-\Delta ^2} } 	
	\left(
	\begin{array}{cc}
		\omega  
		& 
		-\Delta \chi
		\\
		-\Delta \chi
		& 
		\omega 
	\end{array}
	\right) \text{ for } |\omega|>\Delta.
	\label{eq:D_sciam}
\end{align}

Moreover, it is possible to remove the off-diagonal Nambu structures by applying the Bogolyubov-Valatin transformation to the QD degrees of freedom via
\begin{align}
w_{\downarrow}^{\dagger}
&=
\frac{1}{\sqrt{2}}
\left(
d_{\uparrow}^{\vphantom{\dagger}}
+d_{\downarrow}^{\dagger}
\right),
\\
w_{\uparrow}^{\dagger}
&=
\frac{1}{\sqrt{2}}
\left(
-d_{\uparrow}^{\dagger}
+d_{\downarrow}^{\vphantom{\dagger}}
\right),
\label{eq:Bg_Val}
\end{align}
so a scalar hybridization function $D^W(\omega)$ is obtained. The superscript $W$ indicates the use of the new basis for QD electrons~\cite{Zalom-2021}. $D^W(\omega)$ reads
\begin{align}
	D^W(\omega)
	&=
	\Gamma
	\frac{  \mathrm{sgn}(\omega) }{\sqrt{\omega^2-\Delta^2}}
	\left(
	\omega + \Delta \chi
	\right),
	\label{hybrid_WB}
\end{align}
The QD part of the Hamiltonian transforms to
\begin{align}
H_d^W
&=
\sum_{\sigma} 
\varepsilon_{d}
w^{\dagger}_{\sigma}
w^{\vphantom{\dagger}}_{ \sigma}
+U
w^{\dagger}_{\uparrow}
w^{\vphantom{\dagger}}_{ \uparrow}
w^{\dagger}_{\downarrow}
w^{\vphantom{\dagger}}_{ \downarrow},
\nonumber
\\
&
+
\left(\varepsilon_d+U/2\right)
\left(
w^{\dagger}_{\uparrow}
w^{\dagger}_{\downarrow}
+
w^{\vphantom{\dagger}}_{ \downarrow}
w^{\vphantom{\dagger}}_{ \uparrow}
\right).
\label{eq:dotH_W}
\end{align}
In the $W$ basis, the QD Hamiltonian possesses off-diagonal pairing terms proportional to $\varepsilon_d+U/2$, which vanish in the particle-hole symmetric case.

The environment analysis via the hybridization self-energy or its corresponding hybridization function is of both technical and interpretational significance. From the technical point of view, scalar gapped hybridization function \eqref{hybrid_WB} can be directly treated by the log-gap NRG approach of Ref.~\cite{Zalom-2023}. 
On the interpretational level, the scalar nature of the final hybridization function means that the SC-AIM Hamiltonian can be mapped onto a problem where the QD interacts with an insulator-like environment.

In the $W$ basis, the insulator hybridization function possesses a specific particle-hole symmetry when $\varphi=\pi$ as shown in Fig.~\ref{fig:sciam}(d). As $\varphi$ decreases, it then becomes increasingly asymmetric. Connected with the properties of the QD at half-filling, we note that this asymmetry is the only possible source of breaking the dot's doublet GS. In the $W$ basis, the emergence of QPTs is thus rooted in the asymmetry of the hybridization function \cite{Zalom-2021}.

\subsection{Two-terminal QD-based interferometer \label{subsec:interf} }

We now convert the SC-AIM into an interferometer by adding a direct transport pathway between the leads, as shown in Fig.~\ref{fig:ab_ring}(a). This creates an AB ring threaded by the magnetic flux $\Phi_B$. A new term $H_{ij,T}$ is added to obtain the AB-ring Hamiltonian: 
\begin{align}
	H_{AB}
	=
	H_d
	+
	\sum_j H_{j,SC} + H_{j,T}
	+
	\sum_{i\neq j} H_{ij,T}.
	\label{eq:H_AB}
\end{align}
In general, $H_{ij,T}$ is of the form
\begin{align}
	H_{ij,T} 
	&=
	\sum_{\mathbf{k} \mathbf{k}' \sigma} \,
	\left(
	t^*_{\mathbf{k}\mathbf{k}'ij}  
	c^{\dagger}_{\mathbf{k}i\sigma}
	c^{\vphantom{\dagger}}_{\mathbf{k}'j\sigma}
	+ 
	t_{\mathbf{k}\mathbf{k}'ij}  
	c^{\dagger}_{\mathbf{k}'j\sigma}
	c^{\vphantom{\dagger}}_{\mathbf{k}i\sigma}
	\right).
	\label{eq:tunnellead1}
\end{align}
We assume $\mathbf{k}$-independent inter-lead hoppings so that  
\begin{align}
	H_{ij,T} 
	&=
	\frac{1}{N} \sum_{\mathbf{k} \mathbf{k}' \sigma} \,
	\left(
	t^*_{ij}  
	c^{\dagger}_{\mathbf{k}i\sigma}
	c^{\vphantom{\dagger}}_{\mathbf{k}'j\sigma}
	+ 
	t_{ij}  
	c^{\dagger}_{\mathbf{k}'j\sigma}
	c^{\vphantom{\dagger}}_{\mathbf{k}i\sigma}
	\right).
	\label{eq:tunnellead}
\end{align}
In the presence of the magnetic flux penetrating the loop formed by the two arms of this interferometric device, $t_{ij}$ is complex-valued with the condition $t_{ij}=t_{ji}^*$ imposed by definition. The corresponding amplitude and phase can be defined as $t_{ij} = |t_{ij}| e^{-i \phi_{t,ij}/2}$. 

With only two terminals, as considered for the two-terminal system of Fig.~\ref{fig:ab_ring}(a), we drop terminal indices to simplify the general notation to $t_{12} = t e^{-i \phi_t/2}$. Altogether, five phase parameters are present in the corresponding Hamiltonian. Exploiting the gauge invariance, we perform the unitary transformations
$
c^{\dagger}_{\mathbf{k}j\sigma} \rightarrow c^{\dagger}_{\mathbf{k}j\sigma} e^{i \phi_{V,j}/2}
$
which leaves us with a real-valued hopping amplitude $|V_{\mathbf{k}j}|$ but shift the direct hopping amplitude as 
$
t e^{-i \phi_{t}/2} 
\rightarrow
t e^{-i (\phi_{t} - \phi_{V,2} + \phi_{V,1}  )/2 } 
$. 
The BCS gap parameters change in exactly the same way as for SC-AIM discussed previously. Consequently, two gauge-invariant phases become relevant for the interferometric device. They read:
\begin{align}
\varphi 
&\equiv 
\varphi_2 -\varphi_1 + \phi_{V,2} - \phi_{V,1},
\\
\varphi_t 
&\equiv
\phi_t - \phi_{V,2} +\phi_{V,1}.
\end{align}

\section{Analytic properties of the AB-ring interferometer \label{sec:analytic}}

The QD coupling to its environment is fully captured by the discontinuities of the corresponding hybridization self-energy $\mathbb{\Sigma}(z)$. This function can be derived analytically, enabling non-trivial insights. The solution follows the roadmap shown in Fig.~\ref{fig:overview}. We begin with the EOM techniques applied to the general
$n$-terminal interferometric device (Sec.~\ref{subsec:eom}). For the AB-ring with two SC terminals, we then isolate continuous and pole singularities from $\mathbb{\Sigma}(z)$  (Sec.~\ref{subsec:self_energy_2}). The geometric factor $\bm{\chi}$ is introduced as a key quantity characterizing the continuous part in~\ref{subsec:chi}. The isolated poles are shown to give rise to  a side-coupled mode in Sec.~\ref{subsec:side_mode}. We then finally apply the Bogolyubov-Valatin transformations (Sec.~\ref{subsec:bogoliubov_valatin}) to map the interferometric problem onto a single chain insulator-representation.

\subsection{EOM techniques \label{subsec:eom}}

We introduce the Nambu formalism via $D^{\dagger} =  \left(d^{\dagger}_{\uparrow}, 
d^{\vphantom{\dagger}}_{\downarrow} \right)$ and $C_{\mathbf{k}j}^{\dagger} =  \left(c^{\dagger}_{\mathbf{k}j\uparrow}, 
c^{\vphantom{\dagger}}_{-\mathbf{k}j\downarrow} \right)$, so the  Hamiltonian of the general $n$-terminal problem becomes
\begin{align}
 	H_d
 	&=
 	D^{\dagger}
 	\mathbb{E}_d
 	D
 	+
 	U
 	d^{\dagger}_{\uparrow}
 	d^{\vphantom{\dagger}}_{ \uparrow}
 	d^{\dagger}_{\downarrow}
 	d^{\vphantom{\dagger}}_{ \downarrow},
 	\label{eq:dotHnambu}
 	\\
 	H_{j,SC}
 	&=
 	\sum_{\mathbf{k}}
 	C_{\mathbf{k}j}^{\dagger}
 	\mathbb{E}_{\mathbf{k}j}
 	C_{\mathbf{k}j}^{\vphantom{\dagger}},
 	\label{eq:bcsHnambu}
 	\\
 	H_{j,T} 
 	&=
 	\sum_{\mathbf{k}}
 	D^{\dagger}
 	\mathbb{V}_{j}
 	C_{\mathbf{k}j}
 	+
 	\textit{H.c.},
 	\label{eq:tunnelHnambu}
 	\\
 	H_{ij,T} 
 	&=
 	\sum_{\mathbf{k} \mathbf{k}' \sigma} \,
 	C^{\dagger}_{\mathbf{k}i}
 	\mathbb{T}_{ij}
 	C_{\mathbf{k}'j}
 	+
 	\textit{H.c.}
 	\label{eq:directNambu}
\end{align}
with
\begin{align}
	\mathbb{E}_d
	&=
	\varepsilon_d \sigma_z,
	\label{eq:EdNambu}
	\\
	\mathbb{E}_{\mathbf{k}j}
	&=
	\Delta \cos \varphi_j \sigma_x
	-
	\Delta \sin \varphi_j \sigma_y
	+
	\varepsilon_{\mathbf{k}} \sigma_z,
	\label{eq:EkNambu}
	\\
	\mathbb{V}_{j}
	&=
	V_{j} \frac{\sigma_z-\mathbb{1}}{2}
	+
	V^*_{j} \frac{\sigma_z+\mathbb{1}}{2},	
	\label{eq:VNambu}
	\\
	\mathbb{T}_{ij}
	&=
	t_{ij} 
	\frac{\sigma_z-\mathbb{1}}{2}
	+
	t^*_{ij} 
	\frac{\sigma_z+\mathbb{1}}{2},	
	\label{eq:TNambu}
\end{align}
where $\sigma_i$ with $i\in \{ x,y,z\}$ are the Pauli matrices. The property  $t_{ij}=t_{ji}^*$ ensures that $\mathbb{T}_{ij}^{\vphantom{\dagger}}=\mathbb{T}_{ji}^{\dagger}$. 

The total non-interacting Hamiltonian ($U=0$) can be expressed in a compact matrix form as
\begin{align}
	H_0 
	= 
	\Psi^{\dagger} 
	\mathbb{H}_0 
	\Psi^{\vphantom{\dagger}},
	\label{eq:H0}
\end{align} 
where $\mathbb{H}_0$ is the block matrix
\begin{align}
	\mathbb{H}_0
	&=
	\left(
	\begin{array}{c c c c c c}
		\mathbb{E}_{d} & \mathbb{V}_{1} & \mathbb{V}_{2} & \ldots &  \mathbb{V}_{n}
		\\
		\mathbb{V}_{1}^{\dagger} & \mathbb{E}_{\mathbf{k}1} & \mathbb{T}_{12} & \ldots &  \mathbb{T}_{1n}
		\\
		\mathbb{V}_{2}^{\dagger} & \mathbb{T}_{21} & \mathbb{E}_{\mathbf{k}2} & \ldots &  \mathbb{T}_{2n}
		\\
		\vdots & \vdots & \vdots &  \ldots & \vdots 
		\\ 
		\mathbb{V}_{n}^{\dagger} & \mathbb{T}_{n1} & \mathbb{T}_{n2} & \ldots &  \mathbb{E}_{\mathbf{k}n}
	\end{array}
	\right)
	\label{eq:H0matrix}
\end{align}
and $\Psi$ is the composite spinor defined as
\begin{align}
	\Psi^{\dagger}
	=
	\left(
	D^{\dagger}, C_{\mathbf{k}1}^{\dagger}, \ldots, C_{\mathbf{k}n}^{\dagger}
	\right).
	\label{eq:psi_spinor}
\end{align}
It includes the spinors $C_{\mathbf{k}j}^{\dagger}$ for all quasi-momenta $\mathbf{k}$ of lead electrons. We can now derive the hybridization self-energy using the EOM technique. The non-interacting Green's function $\mathbb{G}_0$ is defined through
\begin{align}
	\mathbb{G}_0 
	\left(z \mathbb{1}-\mathbb{H}_0\right)
	=
	\mathbb{1},
	\label{eq:G0_solution}
\end{align}
where $z$ is the complex frequency and $\mathbb{1}$ the unit matrix of the appropriate size. $\mathbb{G}_0$ has the following block structure in terms of $2\times 2$ Nambu matrices:
\begin{align}
\mathbb{G}_0
=
\left(
\begin{array}{c c c c c}
	   \mathbb{G}_{d,0} 
	&  \mathbb{F}^{\dagger}_{\mathbf{k}1} 
	&  \mathbb{F}^{\dagger}_{\mathbf{k}2} 
	&  \ldots 
	&  \mathbb{F}^{\dagger}_{\mathbf{k}n}
	\\ 
	   \mathbb{F}_{\mathbf{k}1}
	&  \mathbb{G}_{c,\mathbf{k}11} 
	&  \mathbb{G}_{c,\mathbf{k}12} 
	&  \ldots 
	&  \mathbb{G}_{c,\mathbf{k}1n}
	\\ 
	   \mathbb{F}_{\mathbf{k}2} 
	&  \mathbb{G}_{c,\mathbf{k}21} 
	&  \mathbb{G}_{c,\mathbf{k}22} 
	&  \ldots 
	&  \mathbb{G}_{c,\mathbf{k}2n}
	\\ 
	   \vdots 
	&  \vdots 
	&  \vdots 
	&  \ldots 
	&  \vdots 
	\\ \mathbb{F}_{\mathbf{k}n} 
	&  \mathbb{G}_{c,\mathbf{k}n1} 
	&  \mathbb{G}_{c,\mathbf{k}n2} 
	&  \ldots 
	&  \mathbb{G}_{c,\mathbf{k}nn}
\end{array}
\right).
\label{eq:G0matrix}
\end{align}
The dependence of all elements on $z$ will be omitted for better readability.

Eq.~\eqref{eq:G0_solution} represents a set of coupled equations for the dot Green's function $\mathbb{G}_{d,0}$ and mixed propagators $\mathbb{F}_{\mathbf{k}i}$:
\begin{align}
\mathbb{G}_{d,0}
\left[
\left(z \mathbb{1}-\mathbb{E}_d\right)
-
\mathbb{G}_{d,0}^{-1}
\sum_{\mathbf{k}} \sum_{i=1}^n 
\mathbb{F}_{\mathbf{k}i}^{\dagger}
\mathbb{V}_{i}^{\dagger}
\right]
&=
\mathbb{1},
\label{eq:dotGreen}
\\
-\mathbb{G}_{d,0}
\mathbb{V}_{i}
+
\mathbb{F}_{\mathbf{k}i}^{\dagger} 
\left(
z \mathbb{1}
-
\mathbb{E}_{\mathbf{k}i} 
\right)
-
\sum_{\mathbf{k}'j\neq i}
\mathbb{F}_{\mathbf{k}'j}^{\dagger}
\mathbb{T}_{j i}
&=
0.
\label{eq:offdiagGreen}
\end{align}
To simplify these expressions, we define
\begin{align}
\mathbb{F}_i^{\dagger}
&=
\sum_{\mathbf{k}}
\mathbb{F}_{\mathbf{k}i}^{\dagger},
\label{eq:F}
\\
\mathbb{G}_i
&=
\sum_{\mathbf{k}}
\frac{\mathbb{1}}{z\mathbb{1}-\mathbb{E}_{\mathbf{k}i}}.
\end{align}
$\mathbb{G}_i$ can be easily evaluated. Assuming a constant normal-state density of states $\rho$ in the leads, one finds
\begin{align}
\mathbb{G}_i
&=
\sum_{\mathbf{k}}
\frac{
	z\mathbb{1} 
	+
	\Delta \cos \varphi_j \sigma_x
	-
	\Delta \sin \varphi_j \sigma_y
	+
	\varepsilon_{\mathbf{k}} \sigma_z}
{z^2-\Delta^2-\varepsilon_{\mathbf{k}}^2}
\nonumber
\\
&=
-
\frac{
	\pi \rho
	\left(z\mathbb{1} 
	+
	\Delta \cos \varphi_j \sigma_x
	-
	\Delta \sin \varphi_j \sigma_y\right)
	}
{\sqrt{\Delta^2-z^2}}
\label{eq:G}
\end{align}

Using \eqref{eq:F} and \eqref{eq:G}, Eq.~\eqref{eq:dotGreen} can be written compactly as
\begin{align}
\mathbb{G}_{d,0}
\left[
\left(z \mathbb{1}-\mathbb{E}_d\right)
-
\mathbb{G}_{d,0}^{-1}
\sum_{i} 
\mathbb{F}_{i}^{\dagger}
\mathbb{V}_{i}^{\dagger}
\right]
&=
\mathbb{1},
\label{eq:dotGreen2}
\end{align}
from which the self-energy $\mathbf{\Sigma}$ is directly read off as
\begin{align}
	\mathbb{\Sigma}(z)
	=
	\mathbb{G}_{d,0}^{-1}
	\sum_{i=1}^n 
	\mathbb{F}_{i}^{\dagger}
	\mathbb{V}_{i}^{\dagger}.
	\label{eq:self-energy}
\end{align}

To obtain a compact form for Eq.~\eqref{eq:offdiagGreen} we first eliminate the $\mathbf{k}$-dependence and obtain a closed set of equations for $\mathbb{F}_j^{\dagger}$. We multiply \eqref{eq:offdiagGreen} by $\left(z\mathbb{1}-\mathbb{E}_{\mathbf{k}i}\right)^{-1}$ and sum over $\mathbf{k}$. Using Eqs.~\eqref{eq:F} and \eqref{eq:G}, this yields

\begin{figure}[H]
	\centering
	\includegraphics[width=0.765\columnwidth]{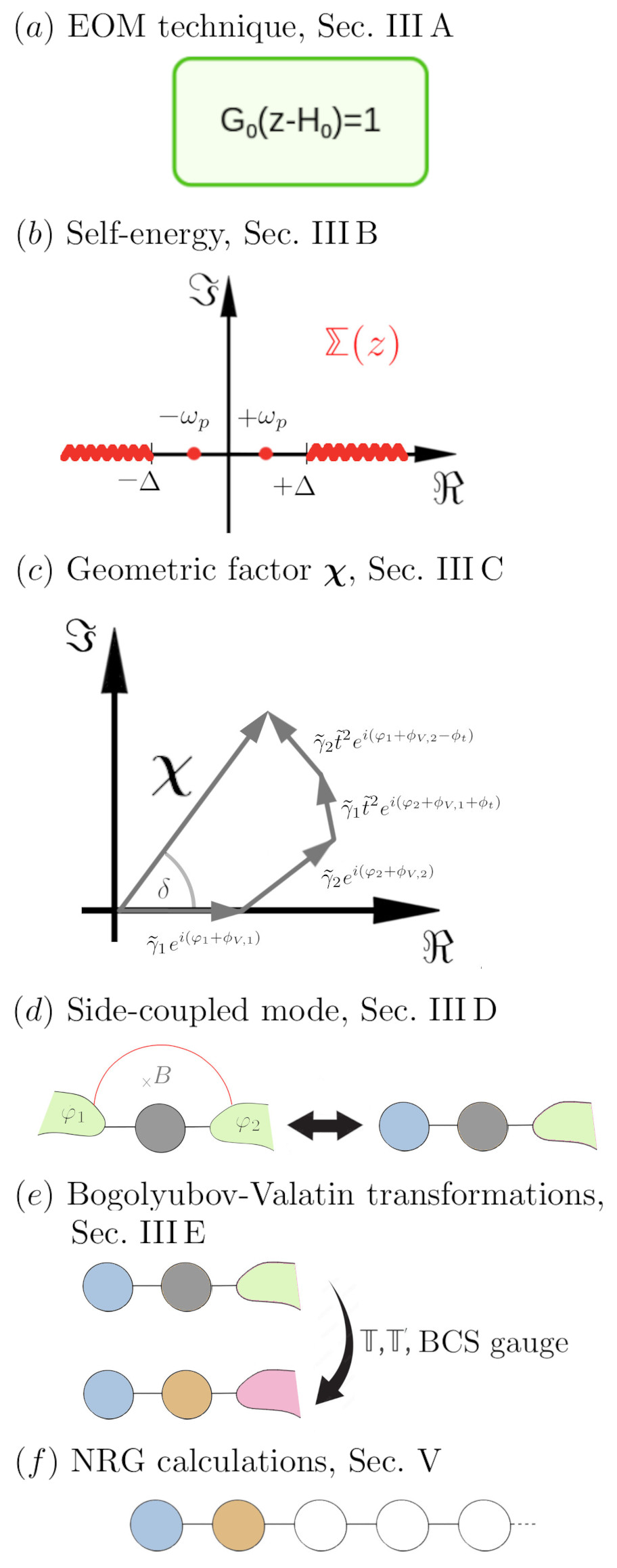}
	\caption{
		$(a)$ 
		Subsection~\ref{subsec:eom} devises the EOM technique for the interferometric problem.
		$(b)$
		\ref{subsec:self_energy_2} demonstrates that the resulting self-energy $\mathbb{\Sigma}(z)$ exhibits diverse singularities in the complex plane. 
		$(c)$
		\ref{subsec:chi} defines geometric factor $\bm{\chi}$ consisting of four complex-valued contributions 
		$\gamma_1 
		e^{i (\varphi_1+\phi_{V,1})}$, 
		$\gamma_1 
		\tilde{t}^2
		e^{i \left(\varphi_2 + \phi_{V,1} + \phi_t \right)}$,
		$\gamma_2
		e^{i (\varphi_2 + \phi_{V,2})}$
		and
		$\gamma_2
		\tilde{t}^2
		e^{i \left(\varphi_1 + \phi_{V,2} - \phi_t \right) }$, which can be rotated counter-clockwise via $\varphi_j \rightarrow \varphi_j-\delta$ due to the gauge freedom. Consequently, real-valued $\bm{\chi}$ can be selected.	
		$(d)$
		In \ref{subsec:side_mode}, the isolated poles of self-energy are reinterpreted as a side-coupled mode of the interacting QD (blue).
		$(e)$ In~\ref{subsec:bogoliubov_valatin} we employ real-valued $\bm{\chi}$ along with the Bogolyubov-Valatin transformations to map the original problem with two SC terminals (green) onto a one-terminal isolator (magenta) problem.		
		$(f)$
		Sec.~\ref{sec:nrg} implements the log-gap NRG to $(e)$ by attaching original QD (orange) to one mode (blue)  and a Wilson chain of non-interacting QDs (white).
		\label{fig:overview}}
\end{figure}

\begin{align}
-
\mathbb{G}_{d,0}
\mathbb{V}_{i}
\mathbb{G}_i
+
\mathbb{F}_{i}^{\dagger} 
-
\sum_{j\neq i}
\mathbb{F}_{j}^{\dagger}
\mathbb{T}_{j i}
\mathbb{G}_i
&=
0.
\label{eq:offdiagGreen2}
\end{align}
This can be expressed as
\begin{align}
\sum_{j=1}^n
\mathcal{F}_{ij}^{\dagger}
=
-\mathbb{G}_{d,0} \mathbb{V}_i \mathbb{G}_i,
\label{eq:Fmatrix}
\end{align}
where $\mathcal{F}^{\dagger}$ is a matrix composed of two by two blocks $\mathcal{F}_{ij}^{\dagger}$ defined as $
\mathcal{F}_{ij}^{\dagger}
=
-\mathbb{F}_{i}^{\dagger} \delta_{ij}
+
\mathbb{F}_{j}^{\dagger}
\mathbb{T}_{j i}
\mathbb{G}_i
$. Most importantly, \eqref{eq:Fmatrix} is linear in $\mathbb{F}_j$ with a right hand side proportional to $\mathbb{G}_{d,0}$. This key property ensures that $\mathbb{F}_j^{\dagger} \propto \mathbb{G}_{d,0}$, which upon insertion into \eqref{eq:self-energy} leads to the cancellation of the $\mathbb{G}_{d,0}^{-1}$ factor in the self-energy.

The off-diagonal blocks of the $\mathcal{F}^{\dagger}$ matrix are only non-zero when a direct hopping between the leads $i$ and $j$ is present. As an important example, let us consider the $n$-terminal cyclic configuration depicted in Fig.~\ref{fig:ab_ring}(b). In such a case, each lead couples only to its nearest neighbor leading to
\begin{align}
	\mathcal{F}^{\dagger}
	=
	\left(
	\begin{array}{c c c c c c}
		  -\mathbb{F}_1^{\dagger} 
		&  \mathbb{F}_2^{\dagger} \mathbb{T}_{21} \mathbb{G}_1 
		&  0 
		&  \ldots 
		&  \mathbb{F}_n^{\dagger} \mathbb{T}_{n1} \mathbb{G}_1
		\\
		   \mathbb{F}_1^{\dagger} \mathbb{T}_{12} \mathbb{G}_2 
		& -\mathbb{F}_2^{\dagger}  
		&  \mathbb{F}_3^{\dagger} \mathbb{T}_{32} \mathbb{G}_2 
		&  \ldots 
		&  0
		\\	
		   0 
		&  \mathbb{F}_2^{\dagger}\mathbb{T}_{23} \mathbb{G}_3 
		& -\mathbb{F}_3^{\dagger} 
		&  \ldots 
		&  0	
		\\
		   \vdots 
		&  \vdots 
		&  \vdots 
		&  \ldots 
		&  \vdots 
		\\ 
		   \mathbb{F}_1^{\dagger} \mathbb{T}_{1n} \mathbb{G}_n 
		&  0 
		&  0 
		&  \ldots 
		& -\mathbb{F}_n^{\dagger}
	\end{array}
	\right).
	\nonumber
	\\
	\label{eq:FmatrixCyclic}
\end{align}
In general, $\mathcal{F}^{\dagger}$ is an $n \times n$ matrix of linearly independent rows with respect to the $2 \times 2$ Nambu matrices $\mathbb{F}_j^{\dagger}$. Consequently, it can be brought into a triangular form from which the corresponding expression for the quantity $\mathbb{F}_n^{\dagger}$ can be directly read off. For cyclic configurations, as presented by Eq.~\eqref{eq:FmatrixCyclic}, the remaining quantities $\mathbb{F}_j^{\dagger}$ with $j<n$ are then obtained by permuting all lead indices in a cyclic manner. In the most general case, the solution of $\mathbb{F}_n^{\dagger}$ needs to be inserted into the $(n-1)$-th line of the triangular form to obtain $\mathbb{F}_{n-1}^{\dagger}$. Repeating the procedure iteratively determines all of the remaining quantities $\mathbb{F}_{j}^{\dagger}$ with $j<n-2$.  

As follows from Eq.~\eqref{eq:self-energy}, the $n$ quantities $\mathbb{F}_j^{\dagger}$ fully define the self-energy $\mathbb{\Sigma}(z)$. By the spectral theorem, $\mathbb{\Sigma}(z)$ is fully determined by its behavior at the real axis of the complex plane, in particular by its singularities. The SC gap leads to the emergence of discontinuity lines along the real axis which terminate at $\pm \Delta$ as shown in Fig.~\ref{fig:overview}(b). They are prescribed by the multi-valued functions appearing in $\mathbb{\Sigma}(z)$. Furthermore, isolated poles can be present in the single-valued as well as multi-valued components of $\mathbb{\Sigma}(z)$. The splitting of $\mathbb{\Sigma}(z)$ into multi-valued (subscript $mv$) and single-valued (subscript $sv$) parts, $\Sigma_{ij}(z)=\Sigma_{ij}^{mv}(z)+\Sigma_{ij}^{sv}(z)$, helps analyzing the different types of singularities. 

\subsection{Self-energy for the two-terminal AB ring \label{subsec:self_energy_2}}

To obtain the self-energy for the two-terminal AB ring, we specialize the general formalism to $n=2$. Equation \eqref{eq:Fmatrix} can be brought into triangular form in a straightforward way. This yields
\begin{align}
	\mathbb{F}_{2}^{\dagger}
	=
	\mathbb{G}_{d,0}
	\left(
	\mathbb{V}_{2}
	\mathbb{G}_{2}
	+
	\mathbb{V}_{1}
	\mathbb{G}_{1}
	\mathbb{T}_{12}
	\mathbb{G}_{2}
	\right)
	\frac{1}{\mathbb{1}
		-
		\mathbb{T}_{21}
		\mathbb{G}_{1}
		\mathbb{T}_{12}
		\mathbb{G}_{2}}.
	\label{eq:F2}
\end{align}
$\mathbb{F}_{1}^{\dagger}$ is obtained by the permutation $1 \leftrightarrow 2$. To obtain the final expression for $\mathbb{\Sigma}(z)$ we insert $\mathbb{F}^{\dagger}_{1}$ and  $\mathbb{F}^{\dagger}_{2}$ into \eqref{eq:self-energy}. This requires explicit parametrization of the phase dependencies entering $\mathbb{G}_j$, $\mathbb{V}_j$ and $\mathbb{T}_{ij}$. 

In the leads, the BCS phases $\varphi_j$ appear in the off-diagonal terms of $\mathbb{G}_{j}$ as 
\begin{align}
	\mathbb{G}_{j}
	&=
	-\frac{\pi \rho}{\sqrt{\Delta^2-z^2}}
	\left(
	\begin{array}{cc}
		  z 
		& \Delta e^{i\varphi_j} 
		\\
		  \Delta e^{-i\varphi_j} 
		& z
	\end{array}
	\right),
\end{align}
while the magnetic flux threading the AB ring introduces a phase in the inter-lead hopping parameter $t_{12} = t_{21}^* \equiv t e^{i\phi_t/2}$ giving
\begin{align}
	\mathbb{T}_{12}
	&=
	\mathbb{T}_{21}^{\dagger}
	=
	t
	\left(
	\begin{array}{cc}
		  e^{i\frac{\phi_t}{2}}  
		& 0 
		\\
		  0
		& -e^{-i\frac{\phi_t}{2}}
	\end{array}
	\right).
	\label{eq:matrixT}
\end{align}
Finally, the dot-lead couplings carry phases $\phi_{V,j}$ and can be parametrized in terms of hybridization strengths $\Gamma_j \equiv \pi \rho |V_j|^2/2$ as
\begin{align}
	\mathbb{V}_{j}
	&=
	\sqrt{\frac{2\Gamma_j}{\pi\rho}} \,
	\left(
	\begin{array}{cc}
		e^{i\frac{\phi_{V,j}}{2}}  & 0 \\
		0 & -e^{-i\frac{\phi_{V,j}}{2}}
	\end{array}
	\right),
	\label{eq:matrixV}
\end{align}
Substituting all parametrizations into \eqref{eq:self-energy} reveals that the self-energy naturally decomposes into two parts:
\begin{align}
\mathbb{\Sigma}(z)
	&=
\mathbb{\Sigma}^{\mathrm{sv}}(z)
+
\mathbb{\Sigma}^{\mathrm{mv}}(z).
\end{align}
with the single-valued part$\mathbb{\Sigma}^{\mathrm{sv}}(z)$ being purely diagonal:
\begin{align}
	\mathbb{\Sigma}^{\mathrm{sv}}(z)
	&=
	\frac{\Gamma^{\mathrm{sv}}}{s(z)}
	\left[
	\begin{array}{cc}
		-s^{\mathrm{sv}}(z)
		& 
		0
		\\
		0
		& 
		s^{\mathrm{sv}}(z)
	\end{array}
	\right],
	\label{eq:sigma_sv}
\end{align}
while the multi-valued contribution $\mathbb{\Sigma}^{\mathrm{mv}}(z)$ reads
\begin{align}
	\mathbb{\Sigma}^{\mathrm{mv}}(z)
	&=
	\Gamma^{\mathrm{mv}} \frac{ \sqrt{\Delta^2-z^2}}{s(z)}
	\left[
	\begin{array}{cc}
		z 
		&
		-\Delta  \bm{\chi}
		\\
		-\Delta \bm{\chi}^* 
		&
		z
	\end{array}
	\right].
	\label{eq:sigma_mv}
\end{align}
The effective hybridization parameters are defined as
\begin{align}
	\Gamma^{\mathrm{sv}}
	&=
	4\tilde{t} 
	\sqrt{\Gamma_1\Gamma_2},
	\\	
	\Gamma^{\mathrm{mv}}
	&=
	2\left(1+\tilde{t}^2 \right)
	\left(\Gamma_1+\Gamma_2\right)
\end{align}
with the dimensionless direct interlead hopping amplitude
\begin{equation}
    \tilde{t} \equiv \pi \rho |t|.
\end{equation}
The factor $\bm{\chi}$ is defined as
\begin{align}
\bm{\chi}
=
\frac{\tilde{\Gamma}^{\mathrm{mv}} }
{\Gamma^{\mathrm{mv}}}
\label{eq:factor_chi}
\end{align}
with
\begin{align}
	\tilde{\Gamma}^{\mathrm{mv}}
	&=
	2\Gamma_1 
	\left[
	e^{i (\varphi_1+\phi_{V,1})}
	+
	\tilde{t}^2
	e^{i \left(\varphi_2 +\phi_{V,1}+\phi_t\right) }
	\right]
	\nonumber
	\\
	&+
	2\Gamma_2
	\left[ 
	e^{i (\varphi_2+\phi_{V,2})}
	+
	\tilde{t}^2
	e^{i \left(\varphi_1+\phi_{V,2}-\phi_t\right) }
	\right].
\end{align}
This $z$-independent complex-valued factor determines the off-diagonal properties of $\mathbb{\Sigma}(z)$ and will be the subject of a detailed analysis in Sec.~\ref{subsec:chi}. Its geometric properties in the complex plane are closely connected to the gauge freedom of BCS phases~\cite{Zalom-2024}. Crucially, $\bm{\chi}$ controls the off-diagonal part of the hybridization in the present problem.

Finally, the functions $s(z)$ and $s_{sv}(z)$ read
\begin{align}
	s(z)
	&=	
	z^2 \left(\tilde{t}^2+1\right)^2
	-
	\Delta ^2 \left(\tilde{t}^4+1\right)
	-
	2\Delta ^2 \tilde{t}^2 \cos
	\left(\varphi +\varphi_t\right),
	\label{eq:sz}
	\\
	s^{sv}(z)
	&=
	\left[
	z^2
	+
	\tilde{t}^2 (z^2-\Delta^2)
	\right]
	\cos{\left(\varphi_t/2\right)}
	-
	\Delta^2 \cos{\left(\varphi_t/2 +\varphi \right)}
	\label{eq:ssvz}
\end{align}
and control the pole singularities. 

Starting with isolated singularities, we note that $\mathbb{\Sigma}^{\mathrm{sv}}(z)$ and  $\mathbb{\Sigma}^{\mathrm{mv}}(z)$ are both proportional to $1/s(z)$. The function $s(z)$ has roots at 
\begin{align}
	\pm \omega_{\mathrm{pole}}
	=
	\pm \Delta
		\sqrt{ 1 - \frac{4\tilde{t}^2}{(\tilde{t}^2+1)^2} \sin^2\left(\frac{\varphi+\varphi_t}{2}\right)}.
	\label{eq:poles}
\end{align}
Both roots lie within the gap region ($|\omega_{\mathrm{pole}}| \leq \Delta$). While in the multi-valued part $\mathbb{\Sigma}^{\mathrm{mv}}(z)$ the nominator cannot cancel the roots of the denominator, in the single-valued part $\mathbb{\Sigma}^{\mathrm{sv}}(z)$ the nominator is proportional to $s^{sv}(z)$ which is another second order polynomial in $z$. It has roots at
\begin{align}
	\omega_p^{\prime} = \pm \Delta
	\sqrt{
		\frac{ 
			\tilde{t}^2 
			+
			\frac
			{ \cos \left(\frac{\varphi_t}{2}  + \varphi \right) }
			{ \cos \left( \frac{\varphi_t}{2} \right) } 
		}
		{ \tilde{t}^2+1 }
	},
\end{align}
which are generally distinct from $\pm \omega_{\mathrm{pole}}$. 

Thus, overall, $\mathbb{\Sigma}(z)$ develops simple poles at $\pm \omega_{\mathrm{pole}}$ whose values are governed solely by $s(z)$. Branch cuts appear only in $\mathbb{\Sigma}^{\mathrm{mv}}(z)$ via the square root function $\sqrt{\Delta^2-z^2}$ and run along the real axis from $-\infty$ to $-\Delta$ and from $\Delta$ to $+\infty$. Point and line singularities together define then the full analytic structure of $\mathbb{\Sigma}(z)$ in the complex plane. 

\subsection{Geometric factor $\bf \bm{\chi}$ \label{subsec:chi} }

Let us now simplify the multi-valued contribution to the hybridization self-energy Eq.~\eqref{eq:sigma_mv}. Here, the factor $\bm{\chi}$ defined in Eq.~\eqref{eq:factor_chi} plays a major role. To show its geometric character, we start by defining the relative hybridization strengths $\tilde{\gamma}_j=2\Gamma_j/\Gamma^{\mathrm{mv}}$ for $j \in \{1,2\}$ and rewrite $\bm{\chi}$ as 
\begin{align}
	\bm{\chi}
	&=
	\tilde{\gamma}_1 
	\left[
	e^{i (\varphi_1+\phi_{V,1})}
	+
	\tilde{t}^2
	e^{i \left(\varphi_2 + \phi_{V,1} + \phi_t \right) }
	\right]
	\nonumber
	\\
	&+
	\tilde{\gamma}_2
	\left[ 
	e^{i (\varphi_2 + \phi_{V,2})}
	+
	\tilde{t}^2
	e^{i \left(\varphi_1 + \phi_{V,2} - \phi_t \right) }
	\right].
	\label{eq:chi}
\end{align}

For some arbitrary choice of BCS phases, Eq.~\eqref{eq:chi} offers a straightforward graphical interpretation for $\bm{\chi}$ via Fig.~\ref{fig:overview}(c). It points in an arbitrary direction of the complex plane, $\arg(\bm{\chi})=\delta$ \cite{Zalom-2024}. By exploiting BCS gauge freedom through simultaneous phase shifts $\varphi_j \rightarrow \varphi_j-\delta$ for both $j \in \{ 1, 2\}$, we can realign $\bm{\chi}$ with the real axis. In this gauge, $\bm{\chi}$ becomes purely real-valued and equals its amplitude $\chi \equiv |\bm{\chi}|$ which can be simplified to
\begin{align}
	\chi^2
	&=
	\frac{\left(\Gamma_1^2+\Gamma_2^2\right) }{(1+t^2)^2(\Gamma_1+\Gamma_2)^2}
	\left[ \tilde{t}^4  + 2\tilde{t}^2  \cos (\varphi_t +\varphi) + 1\right] 
	\nonumber
	\\
	&+
	\frac{2\Gamma_1\Gamma_2 }{(1+\tilde{t}^2)^2(\Gamma_1+\Gamma_2)^2} 
	\, \, \tilde{t}^4  \cos (2 \varphi_t +\varphi) 
	\nonumber
	\\
	&+
	\frac{2 \Gamma_1\Gamma_2}{(1+\tilde{t}^2)^2(\Gamma_1+\Gamma_2)^2}  
	\left[ 2 \tilde{t}^2 \cos (\varphi_t) + \cos( \varphi)\right].
\end{align}	
It satisfies $0\leq \chi \leq 1$. Interestingly, when $\chi=0$, the off-diagonal elements of $\mathbb{\Sigma}^{\mathrm{mv}}$ vanish, marking an important locus of points in the parameter space.

Turning now exclusively to the gauge where $\bm{\chi}$ becomes real, we may finally evaluate the hybridization function $\mathbb{D}_{\mathrm{cont}}(\omega)$ as a jump discontinuity of Eq.~\eqref{eq:sigma_mv}. This yields
 \begin{align}
 	\mathbb{D}_{\mathrm{cont}}(\omega)
 	=
 	\frac{\Gamma^{\mathrm{mv}} \text{sgn}(\omega) \sqrt{\omega ^2-\Delta ^2}}
 	{ s(\omega) } 	
 	\left(
 	\begin{array}{cc}
 		\omega  
 		& 
 		-\Delta \chi
 		\\
 		-\Delta \chi
 		& 
 		\omega 
 	\end{array}
 	\right),
 	\label{eq:D_cont}
 \end{align}
which holds for $|\omega|>|\Delta|$ while elsewhere $\mathbb{D}_{\mathrm{cont}}(\omega)$ is zero. Making $\bm{\chi}$ real-valued is also of a technical advantage as the jump in the imaginary part can be isolated in the square-root prefactor while the rest consists of multiplicative real-valued structures. Consequently, Eq.~\eqref{eq6} can be applied directly in our gauge.

\subsection{Side-coupled mode \label{subsec:side_mode}}

The pole contributions of the hybridization self-energy come from both the single-valued as well as the multi-valued parts of $\mathbb{\Sigma}(z)$. The relevant terms are all proportional to $1/s(z)$ with multiplication factors that are continuous at the pole positions. The discontinuity of $1/s(z)$ across the poles $\pm \omega_{\mathrm{pole}}$ is easily obtained as
\begin{align}
-\frac{\frac{1}{  s(\pm \omega_{\mathrm{pole}}^+)}-\frac{1}{s(\pm \omega_{\mathrm{pole}}^-)}}{2i} 
= 
\frac{\pi }{2\omega_{\mathrm{pole}} (\tilde{t}^2+1)^2 }\delta(\omega \mp \omega_{\mathrm{pole}}).
\label{eq:jump_1/sz}
\end{align}
Consequently, the pole part of the hybridization function, $\mathbb{D}_{\mathrm{pole}}(\omega)$, is 
\begin{align}
	\mathbb{D}_{\mathrm{pole}}(\pm\omega_{\mathrm{pole}})
	&=
	\frac{\pi \Gamma_p}{2}
	\left(
	\begin{array}{cc}
		1 \pm \frac{\varepsilon_p}{\omega_{\mathrm{pole}}}
		& 
		\pm \frac{\Delta_p}{\omega_{\mathrm{pole}}}
		\\
		\pm \frac{\Delta_p}{\omega_{\mathrm{pole}}}
		& 
		1 \mp \frac{\varepsilon_p}{\omega_{\mathrm{pole}}}
	\end{array}
	\right) 
	\delta(\omega \mp \omega_{\mathrm{pole}}).
	\label{eq:pole_hybridization}
\end{align}
with 
\begin{align}
	\Gamma_p
	&=
	\frac{\Gamma^{\mathrm{mv}} } {(\tilde{t}^2+1)^2 } 
	\sqrt{\Delta^2-\omega_{\mathrm{pole}}^2},
	\label{eq:poleG}
	\\
	\varepsilon_p
	&=
	-\frac{ \Gamma^{\mathrm{sv}} s^{sv} (\omega_{\mathrm{pole}})}{  \Gamma^\mathrm{mv} }
	\frac{1}{\sqrt{\Delta^2-\omega_{\mathrm{pole}}^2}},
	\label{eq:poleE}
	\\
	\Delta_p
	&= 
	-\Delta \chi.
	\label{eq:poleD}
\end{align}
A key observation is that $\varepsilon_p$ and $\Delta_p$ fulfill the condition $\varepsilon_p^2+\Delta_p^2=\omega_{\mathrm{pole}}^2$. 

As discussed in Appendix \ref{sec:side_coupled}, the pole part of the hybridization function is of the same form as that generated by a non-interacting QD with energy level $\varepsilon_p$ and induced pairing $\Delta_p$ that is side coupled to the interacting QD via hybridization strength $\Gamma_p$. The Hamiltonian for this side-coupled dot is given as a sum of 
\begin{align}
	H_{\mathrm{side}}
	&=
	S^{\dagger}
	\mathbb{E}_p
	S
	=
	S^{\dagger}
	\left( \Delta_p \sigma_x + \varepsilon_p \sigma_z \right)
	S,
	\label{eq:Hside_nambu}
	\\
	H_{T,s}
	&=
	V_p
	S^{\dagger}
	\sigma_z
	D
	+
	H.c.,
	\label{eq:Htn_side}
\end{align}
where Nambu spinors $S^{\dagger} =  \left(s^{\dagger}_{\uparrow}, s^{\vphantom{\dagger}}_{\downarrow} \right)$ encompass the side-coupled electronic mode created (annihilated) by $s^{\dagger}_{\sigma}$ ($s^{\vphantom{\dagger}}_{\sigma}$) with $\sigma \in \{ \uparrow, \downarrow \}$. Consequently, the $\delta$-peak structures in our hybridization function can be handled as a side-coupled non-interacting dot in the actual NRG calculation in a similar fashion as the zero-frequency peaks appearing in Ref.~\cite{Pruschke-2000-side_coupled}.

\subsection{Bogoliubov-Valatin transformations \label{subsec:bogoliubov_valatin} }

The continuous part of the hybridization function in Eq.~\eqref{eq:D_cont} has the same form as in the ordinary two-terminal SC-AIM reviewed in Sec.~\ref{subsec:two-term}. Following Ref.~\cite{Zalom-2021}, this allows us to diagonalize $\mathbb{D}_{\mathrm{cont}}(\omega)$  through a unitary transformation of the dot degrees of freedom. The  transformation maps the spinor $D^{\dagger}$ to $W^{\dagger} =  \left(w^{\dagger}_{\uparrow}, w^{\vphantom{\dagger}}_{\downarrow} \right)$ as
\begin{align}
	W^{\dagger}
	=
	D^{\dagger}
	\mathbb{T},
\end{align}
where
\begin{align}
	\mathbb{T}
	=
	\sqrt{\frac{1}{2}}
	\left(
	\sigma_x-\sigma_z
	\right).
\end{align}
$\mathbb{T}$ transforms the self-energy $\mathbb{\Sigma}(z)$ unitarily into a diagonal form $\mathbb{\Sigma}^W(z)=\mathbb{T}^{\dagger} \mathbb{\Sigma}(z) \mathbb{T}$ \cite{Zalom-2021} \footnote{The superscript $W$ denotes quantities in the new spinor basis $W$. We omit the corresponding superscript $D$ for the original basis to simplify notation.}.
In the same way, $\mathbb{D}^W_{\mathrm{cont}}(\omega) = \mathbb{T}^{\dagger} \mathbb{D}_{\mathrm{cont}}(\omega) \mathbb{T}$, resulting in
\begin{align}
	\mathbb{D}^W_{\mathrm{cont}}(\omega)
	=
	\frac{
		\Gamma^{\mathrm{mv}}
		\sqrt{\omega ^2-\Delta ^2} \,
	    \text{sgn}(\omega)
	}
	{ s(\omega)  }
	\left(
	\omega  \mathbb{1}
	+ 
	\Delta \chi \sigma_z
	\right)
	\label{eq:continuous_DW}
\end{align}
for $|\omega| \geq \Delta$, while $\mathbb{D}^W_{\mathrm{cont}}(\omega)=0$ in the gap. Due to the diagonal form, the Nambu formalism becomes redundant in the $W$ basis, allowing scalar description. We stress, however, that dropping the Nambu formalism requires different handling of particles and holes \footnote{In the underlying Nambu formalism, the hole propagator $\mathbb{G}_0^W(\omega^+)$ has to be transformed into the corresponding electron propagator. This involves simultaneous change of the frequency sign as well as one extra minus sign for normal ordering of the creation and annihilation operators to obtain the corresponding spin-down electron propagator.}. The scalar hybridization function in the supragap region is thus finally given as
\begin{align}
	D^W_{\mathrm{cont}}(\omega)
	=
	\frac{
		\Gamma^{\mathrm{mv}} 
		\sqrt{\omega ^2-\Delta ^2} \, 
		\text{sgn}(\omega)
	}
	{ s(\omega) }
	\left(
	\omega 
	+ 
	\Delta \chi
	\right).
	\label{eq:continuous_DW_scalar}
\end{align}
$D^W_{\mathrm{cont}}(\omega)$ is always positive which is enforced strictly by the condition $1 \geq \chi \geq 0$. 

\begin{figure*}[ht]
	\includegraphics[width=2.0\columnwidth]{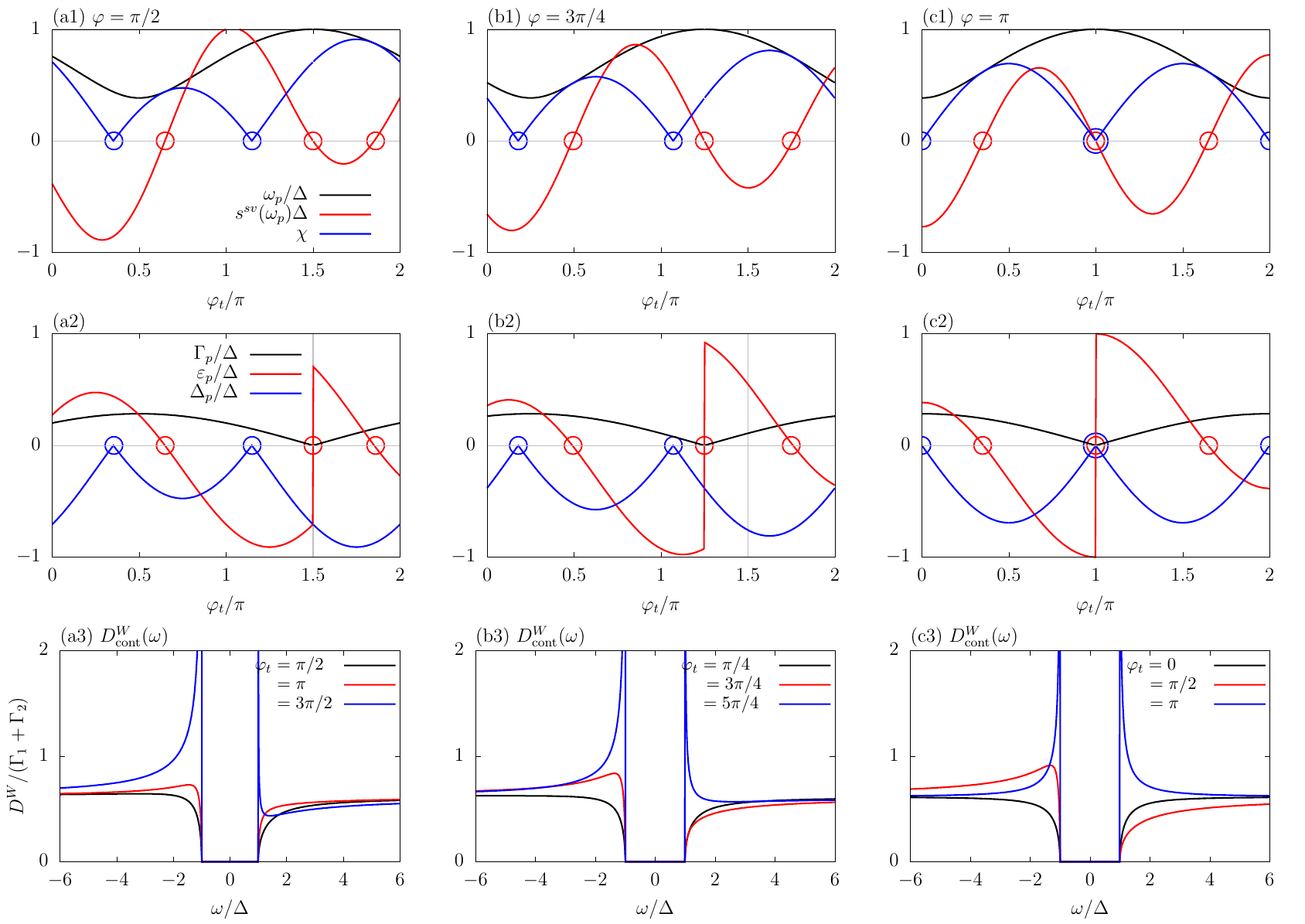}
	\caption{
		(a1)-(c1) Pole position $\omega_{\mathrm{pole}}$, $s^\mathrm{sv} (\omega_{\mathrm{pole}})$ and the geometric factor $\chi$  determine completely the properties of the side-coupled mode. 
		The $\omega_{\mathrm{pole}}$ vs. $\varphi_t$ curves are identical except for the $\varphi$-dependent offsets. 
        $s^{\mathrm{sv}}(\omega_{\mathrm{pole}})$ and $\chi$ contain harmonic functions of various linear combinations of $\varphi$ and $\varphi_t$. Consequently, they change their shape as well as root positions (blue and red circles). 
		(a2)-(c2)
		$\Gamma_p$ depends only on model constants and $\omega_p$, thus it shifts correspondingly but does not change its shape. To accommodate for the condition $\varepsilon_p^2+\Delta_p^2=\omega_{\mathrm{pole}}^2$ both the energy of the side-coupled mode $\varepsilon_p$ as well as the induced pairing $\Delta_p$ change shape and shift the positions of the roots (blue and red circles).
		At $\varphi_t=-\varphi$ the roots of $s^{\mathrm{sv}}(\omega_{\mathrm{pole}})$ and $\chi$ align at the point where $\omega_{\mathrm{pole}}=\Delta$ and the mode decouples due to $\Gamma_p=0$. When additionally $\chi=0$, the AB ring obtains an additional symmetry. This manifests in the $W$ basis as a particle-hole symmetric hybridization function $D^W_{\mathrm{const}}$. In general, $D^W_{\mathrm{const}}$ is particle-hole asymmetric, as shown in panels (a3)-(c3). The particle-hole symmetric $D^W_{\mathrm{const}}$ guarantees a doublet GS which serves as a center of the doublet chimney as in the ordinary SC-AIM shown in Figs.~\ref{fig:sciam}(d1)-(d3).
        Other parameter values are $\Gamma_1/\Delta = 0.25$, $\Gamma_2/\Delta = 0.25$, $\varphi=\pi/2$ and $\tilde{t}/\Delta = 1.5$.
		\label{fig:dqd}}
\end{figure*}

At the gap edges, $D^W_{\mathrm{cont}}(\omega)$ generally exhibits  square-root behavior, see red and black curves in Fig.~\ref{fig:dqd}(a3)-(c3). However, when $\varphi=-\varphi_t$, the prefactor $1/s(z)$ develops poles at $\pm \Delta$. This leads to inverse square root divergencies, see blue curves in Fig.~\ref{fig:dqd}(a3)-(c3), which is the well-known behavior found in the standard SC-AIM~\cite{Zalom-2021}. 

The hybridization function in the $W$ basis consists of two components: a particle-hole symmetric term proportional to $|\omega|\Gamma^{\mathrm{mv}}$ and a particle-hole symmetry-breaking term proportional to $\chi$. Consequently, when $\chi$ vanishes, $\mathbb{D}^W_{\mathrm{cont}}(\omega)$ becomes particle-hole symmetric, as demonstrated in Figs.~\ref{fig:dqd}(a3)-(c3). This is the same mechanism that we had already described for the ordinary SC-AIM in Sec.~\ref{subsec:two-term}.

Since the transformation $\mathbb{T}$ acts on the dot degrees of freedom, we need to transform the dot Hamiltonian \eqref{eq:dotHnambu} accordingly. Following Ref.~\cite{Zalom-2021}, we first decompose it into a quadratic term
\begin{eqnarray}
	H_{\mathrm{dot},0}
	&=&
	D^{\dagger} 
	\left[
	\frac{U}{2} \sigma_x
	+
	\left(\frac{U}{2} + \varepsilon_{d} \right)
	\sigma_z
	\right]
	D^{\vphantom{\dagger}}
	\label{eq:Hdot_d}
\end{eqnarray}
and a mixed quadratic-quartic interaction part that reads
\begin{equation}
	H_U
	=
	U
	d^{\dagger}_{\uparrow}
	d^{\vphantom{\dagger}}_{ \uparrow}
	d^{\dagger}_{\downarrow}
	d^{\vphantom{\dagger}}_{ \downarrow}
	-\frac{U}{2}
	D^{\dagger} 
	\left(
	\sigma_x+\sigma_z
	\right)
	D^{\vphantom{\dagger}}.
	\label{InteractionD}
\end{equation}
Applying transformation $\mathbb{T}$ yields then
\begin{align}
	H_{\mathrm{dot},0}
	&=
	W^{\dagger} 
	\left[
	-\left(\frac{U}{2} + \varepsilon_{d} \right)
	\sigma_x
	-\frac{U}{2}
	\sigma_z
	\right]
	W^{\vphantom{\dagger}}
	\label{eq:Hdot_w}
\\
	H_U^W
	&=
	U
	w^{\dagger}_{\uparrow}
	w^{\vphantom{\dagger}}_{ \uparrow}
	w^{\dagger}_{\downarrow}
	w^{\vphantom{\dagger}}_{ \downarrow}.
	\label{eq:HU_w}
\end{align}
The interaction term $H_U$ thus takes the form of an ordinary Hubbard term in the $W$ basis, with an additional pairing term proportional to $-U/2 - \varepsilon_d$.

The side dot Hamiltonian \eqref{eq:Hside_nambu}, on the other hand,  remains unchanged under $\mathbb{T}$, but the tunneling term \eqref{eq:Htn_side} transforms since it contains the $D$ spinors:
\begin{align}
	H_{T}^{W,S}
	&=
	V_p \,
	S^{\dagger}
	\left(\sigma_z \mathbb{T}\right)
	W
	+
	H.c.
	\nonumber
\\
	&=
	-\frac{V_p}{\sqrt{2}}
	\left(
	s^{\dagger}_{\uparrow}
	w^{\vphantom{\dagger}}_{ \uparrow}
	-
	s^{\dagger}_{\uparrow}
	w^{\dagger}_{ \downarrow}
	+
	s^{\vphantom{\dagger}}_{\downarrow}
	w^{\vphantom{\dagger}}_{ \uparrow}
	+
	s^{\vphantom{\dagger}}_{\downarrow}
	w^{\dagger}_{ \downarrow}
	\right)
	+
	H.c.
\end{align}
with the superscripts $W$ and $S$ indicating the reference against the $W$ and $S$ bases. This new form not only breaks particle number conservation but also spin conservation which is disadvantageous for subsequent NRG calculations. We resolve this by introducing a second unitary transformation $\mathbb{T}'$ that acts on spinor $S^{\dagger}$, producing the spinor $P^{\dagger} =  \left(p^{\dagger}_{\uparrow},p^{\vphantom{\dagger}}_{\downarrow} \right)$:
\begin{align}
	P^{\dagger}
	=
	S^{\dagger}
	\mathbb{T}',
\end{align}
where 
\begin{align}
	\mathbb{T}'
	=
	\sqrt{\frac{1}{2}}
	\left(
	-\sigma_x-\sigma_z
	\right)
\end{align}
is chosen to preserve the form of the tunneling term when both $\mathbb{T}$ and $\mathbb{T}'$ are applied simultaneously. Specifically, $\mathbb{T}'\sigma_z \mathbb{T} = \sigma_z$ ensures that 
\begin{align}
	H_{T}^{W,P}
	&=
	V_{p}
	P^{\dagger}
	\left(\mathbb{T}' \sigma_z \mathbb{T}\right)
	W
	+
	H.c.
	=
	V_{p}
	P^{\dagger}	
	\sigma_z
	W
	+
	H.c.,
	\label{eq:HTp_W}
\end{align}
where the superscripts $W$ and $P$ refer to the spinor bases $W$ and $P$ respectively. Finally, the side mode Hamiltonian transforms as $H^P_{\mathrm{side}} = \mathbb{T}' H_{\mathrm{side}} \mathbb{T}'$, giving
\begin{align}
	H^P_{\mathrm{side}}
	&=
	P^{\dagger}
	\left( \varepsilon_p \sigma_x + \Delta_p\sigma_z \right)
	P.
	\label{eq:Hside_nambu_2}
\end{align}
In the transformed $P$ basis, $\varepsilon_p$ and $\Delta_p$ exchange roles, with $\varepsilon_p$ acting as induced superconductivity and $\Delta_p$ as the energy level.

The reduction of the problem achieved through the procedure detailed in this section is one of the key results of this work, because it identifies the small set of combinations of parameters that control the behavior of the rather complex problem.

\section{Extended symmetry point \label{sec:extended} }

\begin{figure*}[ht]
	\includegraphics[width=2.0\columnwidth]{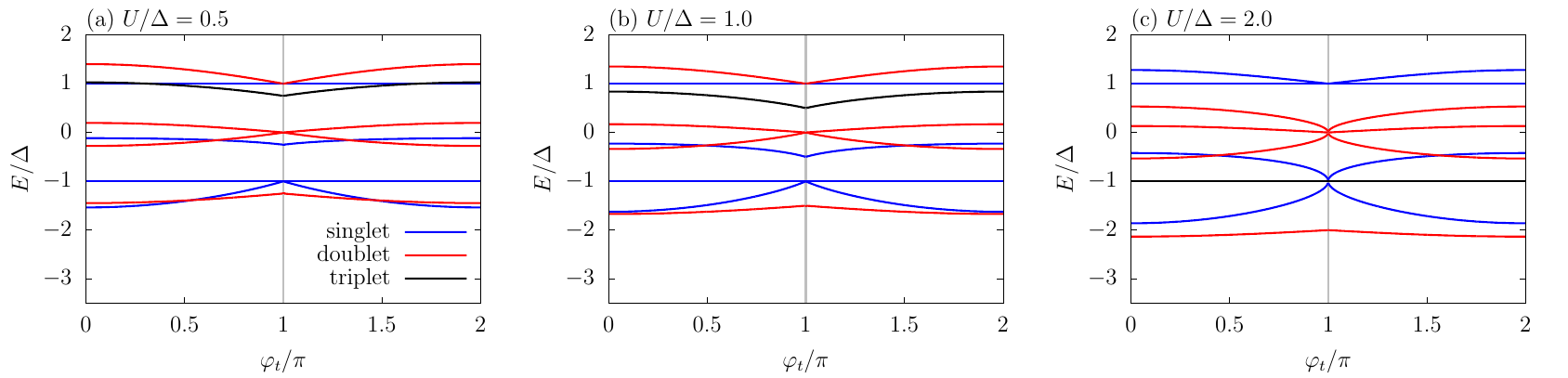}
	\caption{
		Many-body spectra of the DQD system comprised of the interacting QD and the side-coupled mode $p$ for three values of $U$. The model parameters are the same as in Fig.~\ref{fig:dqd} with $\varphi=\pi$. The interacting dot is kept at half-filling, $\varepsilon_d=-U/2$. The vertical gray line denotes the decoupling of the DQD system for $\Gamma_p=0$ which guarantees an overall doublet GS. With varying $\varphi$ the spectra only shift as the decoupling condition moves according to $\varphi_t=-\varphi$. 
		\label{fig:dqd_states}}
\end{figure*}

The interferometer model under study is governed by two phases, $\varphi_t$ and $\varphi$, which are highly tunable in practical applications, while the remaining parameters $\tilde{t}$, hybridization strengths $\Gamma_1$, $\Gamma_2$ and QD parameters $U_d$, $\varepsilon_d$ can only be varied if gate electrodes are available.

The central quantity that allows to intuitively navigate this vast parametric space is the geometric factor $\chi$. The locus of points where this quantity vanishes corresponds to the cases where the hybridization function becomes diagonal in the $D$ basis, which in turn leads to a particle-hole symmetric hybridization function in the $W$ basis. Remarkably, complementing the information from $\chi$ with that from $\omega_{\mathrm{pole}}$ is sufficient to decipher the system's key properties.

To illustrate this idea, we show the symmetric-coupling case with $\Gamma_1/\Delta = \Gamma_2/\Delta = 0.25$ and $\tilde{t}/\Delta = 1.5$. Considering three cases of $\varphi \in \{ \pi/2, 3\pi/4, \pi\}$ we vary the magnetic flux phase $\varphi_t$ and plot the corresponding $\omega_{\mathrm{pole}}$, $s^{\mathrm{sv}}(\omega_{\mathrm{pole}})$, and $\chi$ functions in Figs.~\ref{fig:dqd}(a1)-(c1). Notably, while the shape of $\omega_{\mathrm{pole}}$ vs. $\varphi_t$ remains the same and only its offset depends on $\varphi$, as per Eq.~\eqref{eq:poles}, the two other curves change in shape and phase shift relative to the $\omega_{\mathrm{pole}}$ curve as $\varphi$ changes. 

Using Eqs.\eqref{eq:poleG}-\eqref{eq:poleD}, we can now immediately elucidate  the evolution of the properties of the side-coupled mode. Its hybridization $\Gamma_p$ depends apart from model-specific parameters, only on $\omega_p$. Consequently, $\Gamma_p$ will have the same shape for all values of $\varphi$ with respect to $\varphi_t$, but it will be shifted together with its roots as shown in Figs.~\ref{fig:dqd}(a2)-(c2). In contrast, $\varepsilon_p$ and $\Delta_p$ modulate in response to changes in $s^{\mathrm{sv}}(\omega_{\mathrm{pole}})$ and $\chi$. A striking feature is the discontinuity in $\varepsilon_p$ at $\varphi_t=-\varphi + 2m\pi$, where generally $m \in \mathbb{N}$ with Fig.~\ref{fig:dqd} showing the $m=1$ discontinuity. This abrupt jump occurs because a pole in $s^{\mathrm{sv}}(\omega_{\mathrm{pole}})$ aligns precisely with the condition $\omega_{\mathrm{pole}}=\Delta$ where the side-coupled mode decouples from the system due to $\Gamma_p=0$. These points are marked by overlapping black and red circles in Figs.~\ref{fig:dqd}(a2)-(c2).

Notably, at these points the continuous parts of the hybridization function $D^W_{\mathrm{const}}$ develop BCS-like divergences as highlighted by blue curves in Figs.\ref{fig:dqd}(a3)-(c3). Another distinct situation occurs when $\chi=0$, as marked by blue circles in Figs.~\ref{fig:dqd}(a2)-(c2). Here, the hybridization function $D^W_{\mathrm{const}}$ becomes particle-hole symmetric. When these two special cases occur simultaneously we obtain a highly symmetric scenario reminiscent of the SC-AIM: a particle-hole symmetric hybridization function is coupled to an interacting dot, while the side-coupled mode is actually detached. This guarantees a degenerate ground state.

Let us therefore turn our attention to the many-body spectrum of the system formed by the interacting dot and the side-coupled mode, which we will refer to as the double quantum dot (DQD) in what follows. We keep the same parameters as used in Fig.~\ref{fig:dqd} and first keep $\varphi=\pi$. The many-body energy spectrum can be easily determined as the Hilbert space contains only 16 states. The results are presented in Fig.~\ref{fig:dqd_states} for three values of $U/\Delta \in \{ 0.5, 1.0, 2.0 \}$ with $\varphi=\pi$. 

The decoupling points, $\Gamma_p=0$, occur periodically at $\varphi_t=(2m-1)\pi$. For the selected range of $\varphi_t$, only the $m=1$ point corresponding to decoupling at $\varphi_t=\pi$ is shown in Fig.~\ref{fig:dqd_states} (highlighted by the gray vertical line). Here, a doublet GS is always guaranteed and thus forms the core of the doublet chimney that stretches symmetrically out of that point in the $\varphi_t$ cross-section. At $U/\Delta=1/2$, the interactions pull one of the excited singlets below the doublet GS at around $\varphi_t/\pi=1 \pm 0.45$ (marked by a black circle) resulting in a singlet-doublet QPT. For the two larger values of $U/\Delta$, the doublet GS stretches from the decoupling point into the whole range of $\varphi_t$ values, thus preventing any QPTs driven by $\varphi_t$. This mechanism resembles thus the familiar doublet chimney effect in the standard two-terminal SC-AIM.

While the many-body spectrum of Fig.~\ref{fig:dqd_states} is presented at an arbitrarily chosen $\varphi=\pi$, we stress that varying $\varphi$ only rigidly shifts the entire spectrum to the point where the center of the doublet chimney aligns with the decoupling point of the DQD system at $\varphi_t=-\varphi +2m\pi$ with $m \in \mathbb{N}$. This is due to the shifting properties of $\omega_{\mathrm{pole}}$ and $\Gamma_p$ with respect to $\varphi$ and the condition $\varepsilon_p^2+\Delta_p^2=\omega_{\mathrm{pole}}^2$.

The DQD spectrum thus encapsulates the qualitative features of the doublet chimney. Nevertheless, it cannot properly account for the effects of the continuous hybridization function. This is easily seen by realizing that the DQD system serves as the starting point for the actual NRG calculation, where its eigenstates become ``dressed'' by a Wilson chain generated from the continuous part of the hybridization function. Here, the $\varphi$ dependence enters the problem significantly via the properties of the hybridization functions $\mathbb{D}_\mathrm{cont}$. The mirror symmetry of the DQD spectra around the decoupling point will thus be broken once $\varphi \neq -\pi + 2m\pi$ as the $\chi=0$ condition will not align with the decoupling point of the DQD system. In a situation of $\chi \neq 0$ the many-body spectrum of the decoupled DQD gets dressed by a particle-hole asymmetric hybridization function that breaks the underlying simple symmetry. This is of essential importance especially for the emergence of Josephson diode effect \cite{Souto-2022,Cheng-2023}. 

Thus, one can expect the renormalized spectra to be $\varphi$-dependent with singlet-doublet transition points being strongly renormalized or even  completely missing. Nevertheless, for $\varphi=\pi$ we know that the root of $\chi$ aligns precisely with the root of $\omega_{\mathrm{pole}}$. Consequently, the particle-hole symmetric Wilson chain associated with $\chi=0$ dresses the high-symmetry points of the many-body-spectra of the DQD. The initial doublet GS of the DQD cannot be  renormalized away in this way because of the symmetry, so the GS parity is preserved through-out the whole RG flow. We stress that this extended symmetry point is essentially identical to that of the regular two-terminal SC-AIM as found in Ref.\cite{Zalom-2021}. In both cases, this high-symmetry scenario appears strictly at $\varphi=\pi$ and guarantees a doublet GS at half filling. Consequently, it leads to the phenomenon of the doublet chimney \cite{Pavesic-2024}.

Understanding the formation of doublet ground states for the Aharonov-Bohm interferometer requires exploring the full parameter space of the problem. However, previous work~\cite{Karrasch-2009-ab} did not take the crucial flux-control of the interferometric phase into account, setting $\varphi_t=0$ throughout. Since the core of the doublet chimney is located at $\varepsilon_d=-U/2$, $\chi=0$, and $\Gamma_p=0$, the decoupling point is fixed to $\varphi=-\varphi_t=0$. On the other hand, the coupling symmetric scenario $\Gamma_1=\Gamma_2$ employed in Ref.~\cite{Karrasch-2009-ab} yields the geometric factor $\chi=\cos(\varphi/2)$, which cannot vanish at this particular decoupling point. Consequently, the phase diagrams of Ref.~\cite{Karrasch-2009-ab} did not contain the core of the doublet chimney. Instead, the restriction to $\varphi_t=0$ effectively sliced through the doublet chimney at various locations, giving rise to the observed reentrant singlet-doublet QPTs~\cite{Karrasch-2009-ab}. 

The generalization of the conditions for the occurrence of the doublet chimney to the Aharonov-Bohm interferometer represents one of the key results of this work, complementing purely numerical approaches with qualitative physical insight. Studying the detailed structure of the doublet chimney requires unbiased numeric methods such as the log-gap NRG employed here.

\section{NRG solution \label{sec:nrg} }

\begin{figure*}[t]
	\includegraphics[width=2.1\columnwidth]{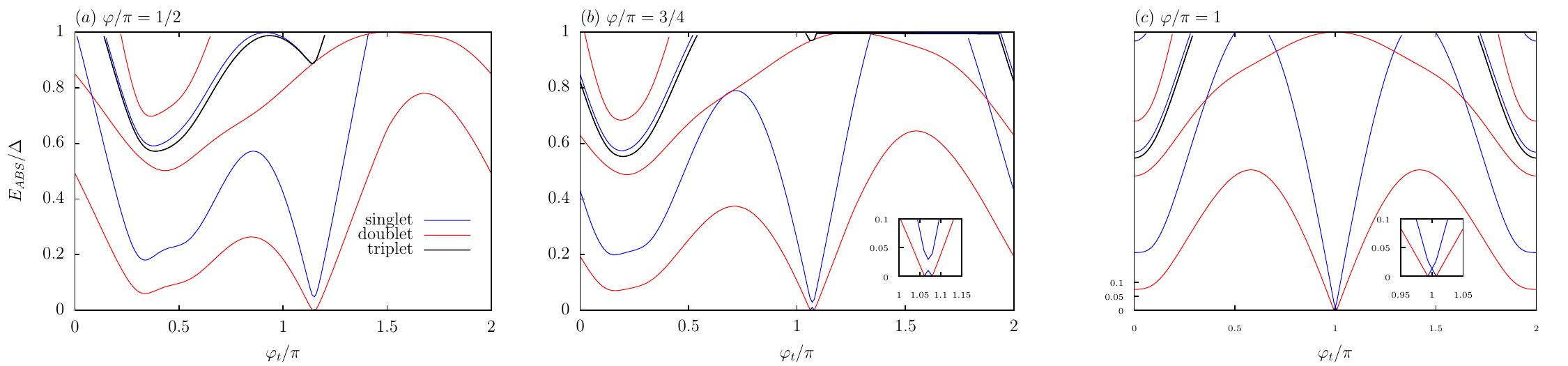}
	\caption{
		Subgap state energies $E_{ABS}$ for $U/\Delta=3$, $\Gamma_1/\Delta=\Gamma_2/\Delta=0.25$, $\tilde{t}=1.5$ for three values of $\varphi/\pi \in \{ 1/2, 3/4, 1\}$. Note that only panels (b) and (c), where $\varphi/\pi \in \{3/4, 1\}$, show singlet-doublet QPTs. These occur in the vicinity of $\chi=0$ solutions. Moreover, in panel (c), the doublet ground state is accompanied by the crossing of excited singlets at $\varphi_t=-\varphi + 2m\pi$ ($m=1$ in the presented plot) as guaranteed by the presence of an extended symmetry herein.
		\label{fig:abses}}
\end{figure*}

This unbiased numerical approach allows us to explore the full parameter space including both $\varphi$ and $\varphi_t$, extending beyond the restricted scenario $\varphi_t=0$ considered in the FRG study of Ref.~\cite{Karrasch-2009-ab}. To this end we dress the bare DQD spectra with Wilson chains corresponding to the continuous hybridization functions.
In order to perform the renormalization procedure, scaling laws need to be identified so that the many-body spectra emerging during the iterative extension of the chain converge to RG fixed points. 
For gapped environments, it has recently been shown that the so-called log-gap discretization of the continuous hybridization function is required for a stable and well-converged RG flow~\cite{Liu-2016,Zalom-2023}. The analytic transformations have brought the present problem into a basis with a scalar hybridization function. Technically, it is therefore of the same complexity as the SC-AIM or the gapped Anderson model as treated in~\cite{Zalom-2023}. 

All results presented in this section have been obtained using NRG Ljubljana in iterative diagonalizations with $\Lambda=2$. At least $2000$ states were kept at every even iteration accommodating the alternating discard scheme of Ref.~\cite{Zalom-2023}. 
We fix $U/\Delta=3$ and $\Gamma_1/\Delta=\Gamma_2/\Delta=0.25$ throughout this section.

\subsection{Subgap spectra at $\bf \tilde{t} = 1.5$ \label{subsec:abses} }

Let us first focus on how the doublet chimney is generated with respect to the interferometric phase $\varphi_t$. To this end we select a moderately large direct hopping amplitude $\tilde{t}=1.5$ at three values of $\varphi/\pi \in \{ 1/2, 3/4, 1 \}$ with the resulting dressed ABS spectra shown in Fig.~\ref{fig:abses}(a)-(c). We observe several key features: i) the spectra are $\varphi$-dependent unlike their DQD precursors from Sec.~\ref{sec:extended}; ii) the subgap spectra are significantly richer compared to the standard SC-AIM and contain spin singlets, doublets, and triplets; iii) the $\varphi=\pi$ case contains an extended symmetry point that serves as the core of a doublet chimney that protrudes up to $\varphi/\pi = 3/4$, as observed in Fig.~\ref{fig:abses}(b); and iv) the singlet-doublet QPTs occur in the AB interferometer for all cases depicted in Fig.~\ref{fig:abses}.

Only the sub-gap spectrum at $\varphi=\pi$ is mirror-symmetric around the decoupling point $\varphi_t=-\varphi$ which is due to the particle-hole symmetry of $D^W(\omega)$ (and of the Wilson chain it generates). The resulting renormalization substantially impacts the initial spectra of the DQD system. For $U/\Delta=3$, these posses an exclusively doublet GS for all values of $\varphi_t$. However, dressing these with the corresponding Wilson chains leaves only a narrow doublet GS regions appearing around $\varphi_t/\pi=-\varphi+2m\pi$ with $m \in \mathbb{N}$ ($m=1$ in Fig.~\ref{fig:abses_t_dep}). Here the root of $\chi$ are localized. Interestingly, in Figs.~\ref{fig:abses_t_dep}(a) and (b) the doublet GS of the initial DQD system at the decoupling point $\Gamma_p=0$ is renormalized away in favor of a singlet GS due to the substantial particle-hole asymmetry. Only for $\varphi=\pi$, a doublet GS is guaranteed because of the alignment of the decoupling point and the root of $\chi$ at $\varphi_t=-\varphi+2m\pi$. Still, the doublet chimney is severely shrunken to a region of approximately $0.995<\varphi_t<1.005$. The respective roles of $U$ and $\tilde{t}$ in this renormalization effect are discussed in Sec.~\ref{subsec:abses_t0}.

In addition to the singlet and doublet subgap states, well known from the ordinary two-terminal SC-AIM, the spin triplet states (red) appear in the upper half of the gap. This happens because the finite inter-lead tunneling $\tilde{t}$ pulls down linear combinations of Bogoliubov modes from both contacts toward the GS, thereby forming new subgap levels (precisely those described by the side-coupled mode). These levels can be either unoccupied, doubly occupied, or occupied by a single particle with either spin orientation. These states mix with the conventional subgap states induced by the QD, leading to complex spectral evolution. This is another important finding of this work.

While no triplet GSs are observed, standard singlet-doublet transitions occur depending on how closely the selected model parameters approach the doublet chimney anchored at $\varphi=\pi$ and $\varphi_t=-\pi+2m\pi$ with $m \in \mathbb{N}$. Similar to the SC-AIM, this high-symmetry point serves as a precursor for the emergence of QPTs and is therefore essential for qualitative understanding of the emergent physics. In the AB ring device, however, both phases $\varphi$ and $\varphi_t$ must be taken into account. Consequently, the geometric factor $\chi$ becomes an important guide for understanding the potential outcomes. However, subtle interplay with the side-coupled mode, as encoded in its frequency $\omega_p$, takes place in the AB ring.

While the DQD model of Sec.~\ref{sec:extended} correctly describes the basic interplay, dressing its eigenstates with particle-hole asymmetric Wilson chains of NRG strongly renormalizes the outcome by favoring singlet GSs. The extent of the doublet chimney in the $\varphi_t$ parameter is thus largely overestimated by the bare DQD system. In the $\varphi$ parameter, the renormalized doublet chimney shows overall a smaller extent compared to the bare DQD model and the artificial shift of spectra with respect to $\varphi$ was lifted. In general, however, the doublet chimney expands considerably along $\varphi$ axis which explains the persistency of doublet phase across various regimes as observed already in Ref.~\cite{Karrasch-2009-ab}. 

\subsection{Subgap spectra as $\bf \tilde{t} \rightarrow 0$ \label{subsec:abses_t0} }

\begin{figure*}[t]
	\includegraphics[width=2.1\columnwidth]{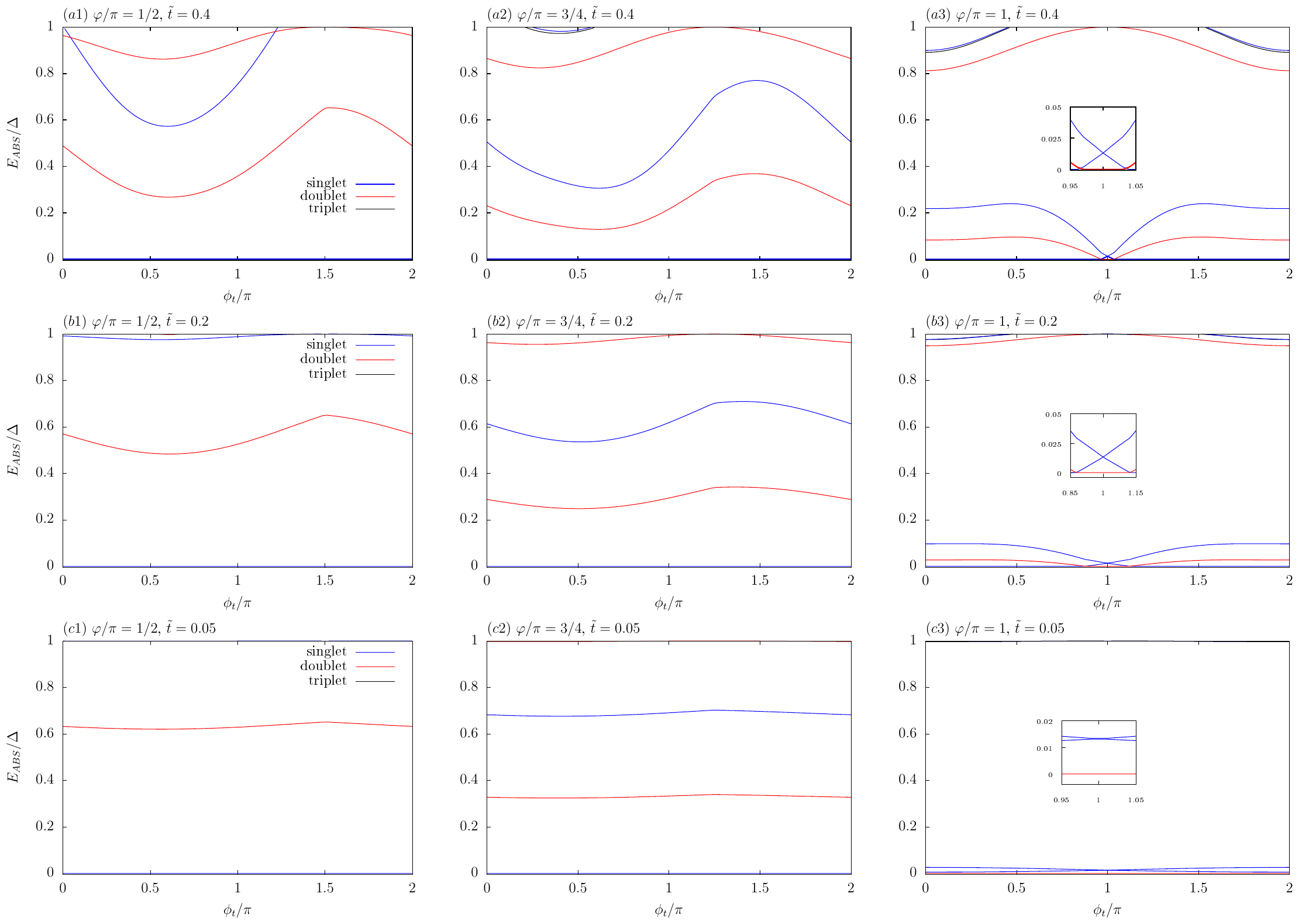}
	\caption{
		$\varphi_t$-dependence of subgap state energies $E_{ABS}$ for $U/\Delta=3$, $\Gamma/\Delta=0.5$  at three values of $\varphi/\pi \in \{ 1/2,3/4,1 \}$ with $\tilde{t}$ decreasing from $0.4$ down to $0.05$. As $\tilde{t} \rightarrow 0$, almost constant ABS energies are obtained indicating transition into the behavior of standard two-terminal SC-AIM at $\tilde{t}=0$.
		\label{fig:abses_t_dep}
        }
\end{figure*}

Having established the role of the interferometric phase $\varphi_t$ in Sec.~\ref{subsec:abses}, we now investigate further how the doublet chimney emanates from its core into the parametric space of the interferometric set-up. The strong renormalization effects on the bare DQD system suggest potential interplay of various model parameters. To identify these, let us discuss the evolution of subgap spectra as $\tilde{t}$ decreases.

To this end we selected the same $U$, $\Gamma_1$, $\Gamma_2$, $\varphi$ as in the previous section and varied $\tilde{t}$ values, see Fig.~\ref{fig:abses_t_dep}. Most notably, the triplet states are largely suppressed already at $\tilde{t}\leq 0.4$ and almost merge with the gap edge. More pronounced $\varphi_t$-dependence is visible only specifically at $\tilde{t}= 0.4$ and $\varphi=\pi$, as shown in Fig.~\ref{fig:abses_t_dep}(a3). 

Consequently, the subgap spectrum is dominated by singlets and doublets for $\tilde{t} \leq 0.4$. Concentrating at the doublet chimney, which is guaranteed to exist at $\varphi=\pi$, let us focus on the insets of Figs.~\ref{fig:abses_t_dep}(a3), (b3) and (c3) as $\tilde{t}$ decreases. While initially, at $\tilde{t}=0.4$, its $\varphi_t$ extent has comparable width as in the $\tilde{t}=1.5$ case, it increases significantly as $\tilde{t} \rightarrow 0$. At $\tilde{t}=0.05$ an exclusive doublet GS covers all $\varphi_t$ values marking thus the maximum possible extent of the chimney in this parameter. 

In the parameter $\varphi$, we observe an opposite dependence. The cases $\varphi/\pi = 1/2$ and $\varphi/\pi = 3/4$ posses exclusively singlet ground states for $\tilde{t} \leq 0.4$ as all $\varphi_t$-dependencies significantly flatten out and become almost constant already at $\tilde{t}=0.05$. This behavior is expected as the AB interferometer moves closer to an ordinary two-terminal SC-AIM behavior with the value of $\varphi_t$ becoming increasingly irrelevant. 

Looking at the sequence of Figs.~\ref{fig:abses_t_dep}(c1)-(c3) we recognize a pattern familiar from the SC-AIM. Starting at $\varphi=\pi$, we observe a doublet GS and two singlets. The two singlets increasingly separate as $\varphi$ decreases and the system undergoes a singlet-double QPT before reaching $\varphi/\pi =3/4$. Consequently, in Fig.~\ref{fig:abses_t_dep}(c2) a singlet GS is followed by an excited doublet with a second singlet shifted towards the gap edge. This outer singlet merges subsequently with the continuum above the gap leaving only a singlet GS and one excited doublet state in the gap, see Fig.~\ref{fig:abses_t_dep}(c1). 

We can therefore conclude that $\tilde{t}$ has the largest impact on the extent of the doublet chimney in the $\varphi_t$ parameter while the impact of $U$ and $\varphi$ follows largely the pattern of the standard SC-AIM. Generally, as $\tilde{t}$ increases a larger departure from SC-AIM occurs with triplet in-gap states moving from the continuous part into the sub-gap region. Here, the doublet chimney effect becomes dominated by the roots of the geometric factor $\chi$.

\subsection{$\bf \varphi$-dependence of subgap spectra \label{subsec:abses_phi}}

\begin{figure*}[t]
	\includegraphics[width=2.1\columnwidth]{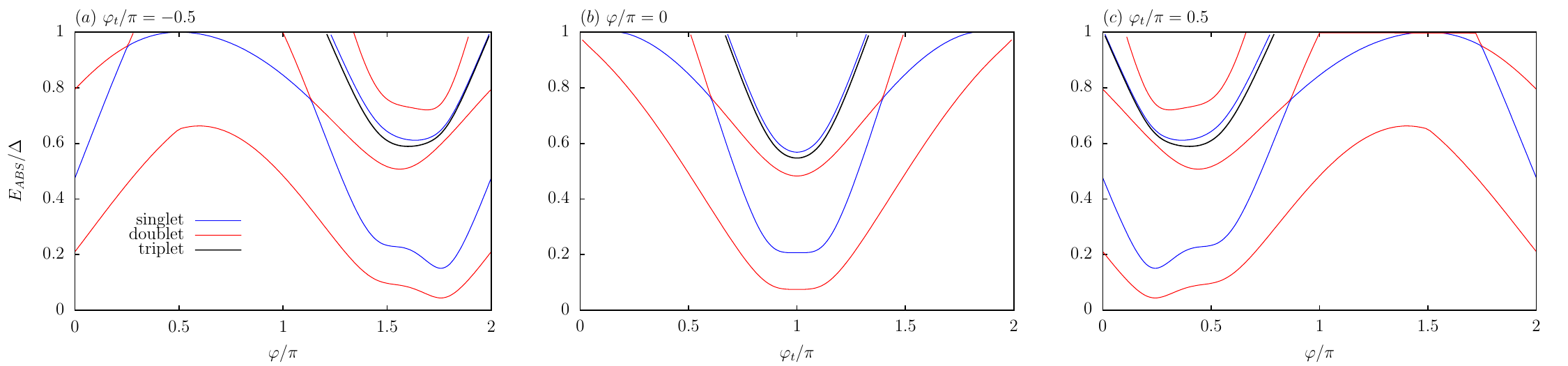}
	\caption{
		Subgap state energies $E_{ABS}$ for $U/\Delta=3$ $\Gamma/\Delta=0.5$ $\tilde{t}=1.5$ at three selected values $\varphi_t/\pi$. in panel (a) $\varphi_t/\pi=-1/2$, in (b)  $\varphi_t/\pi=0$ and in (c) $\varphi_t/\pi=1/2$.
		\label{fig:abses_phidep}}
\end{figure*}

Finally, we investigate the $\varphi$-evolution of subgap spectra at fixed $\varphi_t/\pi \in \{ -1/2,0,1/2 \}$ with the remaining model parameters identical to those used in Sec.~\ref{subsec:abses}. The results are shown in Fig.~\ref{fig:abses_phidep}. The choice of $\varphi_t$ ensures exclusively a singlet GS in all three cases. Once again a complex picture of singlet, doublet and triplet in-gap states emerges. 

The most striking feature illustrated in Fig.~\ref{fig:abses_phidep} is the manifestation of certain symmetries with respect to $\varphi$ parameter. Clearly, at $\varphi_t=0$ the spectrum is mirror-symmetric with respect to the $\varphi/\pi=1$ axis with no analogous symmetry acquired in the other two cases. This aligns exactly with the findings of Ref.~\cite{Karrasch-2009-ab} in the restricted parameter space of $\varphi_t=0$. Additionally, we find that the two presented cases beyond this limitation, namely $\varphi_t/\pi=\pm 1/2$, are related to each other in the following way: changing $\varphi_t \rightarrow -\varphi_t$ amounts to changing $\varphi \rightarrow -\varphi$ to obtain the same spectra.

The $\varphi$ behavior of ABS spectra clearly indicates when Josephson diode effect is present in the system. For instance, the presence of the mirror-symmetry in Fig.~\ref{fig:abses_phidep}(b) means that taking derivatives of many-body spectra with respect to $\varphi$ would result in anti-symmetric functions. Since at zero temperature, Josephson current is proportional to $\varphi$ derivative of ground state energy, it would become anti-symmetric with zero value clearly acquired at $\varphi/\pi=0$. Consequently, positive and negative critical supercurrents would be identical with no Josephson diode effect present. Going beyond the limitation of Ref.~\cite{Karrasch-2009-ab} by setting $\varphi_t \neq 0$, the mirror-symmetry was explicitly lifted and the corresponding $\varphi$ derivatives are no longer anti-symmetric. Consequently, different forward and backward critical supercurrents appear as expected with such an interferometric approach~\cite{Souto-2022}. A full NRG study of the diode effect is, however, beyond the scope of the present paper as explained in App.~\ref{sec:josephson}.

\section{Conclusions \label{sec:conclusions} } 

We have presented a comprehensive analytical and numerical study of an Aharonov-Bohm interferometer with SC terminals and a strongly correlated QD embedded into one of its arms. Through rigorous derivation using equation of motion techniques, we have proven that this double-path interferometer is spectrally equivalent to a simpler system consisting of an interacting quantum dot coupled to a non-interacting side-coupled mode and a semiconductor-like lead.

Our analysis reveals that the interferometer's behavior arises from a competition between the properties of the side-coupled mode and the asymmetry of the semiconductor hybridization function emerging from a unitary transformation of the basis of QD electronic degrees of freedom. The geometric factor $\chi$ plays a central role in this competition, controlling the particle-hole symmetry of the semiconductor hybridization function. When $\chi = 0$, the hybridization becomes particle-hole symmetric in the transformed basis. When this condition coincides with the decoupling of the side-coupled mode ($\Gamma_p = 0$), an extended symmetry point emerges via a mechanism analogous to that in SC-AIM and serves as a core for the doublet chimney. This gives crucial analytic understanding of the re-entrant occurrence of singlet-doublet phase transitions as observed already in the restricted $\varphi_t=0$ scenario of Ref.~\cite{Karrasch-2009-ab}.

The log-gap NRG calculations confirm these analytical insights and demonstrate that the ABS spectrum of the interferometer is significantly richer than that of the standard SC-AIM which is in line with a previous FRG study of Ref.~\cite{Karrasch-2009-ab}. In addition to the familiar singlet and doublet states, triplet states appear due to the finite inter-lead tunneling, which pulls linear combinations of Bogoliubov modes from both contacts toward the GS. Despite this increased complexity, singlet-doublet QPTs persist, and the doublet chimney phenomenon also exists in the interferometric geometry.

Our results show that doublet chimney structure emerges in the phase diagram at $\varphi_t=-\varphi+2m\pi$ ($m \in \mathbb{N}$) and is of crucial significance for the understanding of the phase diagram. As the direct tunneling amplitude $\tilde{t}$ decreases toward zero, the spectrum simplifies and approaches the behavior of the standard SC-AIM, confirming the consistency of our approach.

The equivalence between the complex interferometric device and the simpler effective model provides both conceptual clarity and computational advantages for studying quantum transport in superconducting quantum interferometers. Introduction of a direct transport path enables also the Josephson diode effect which appears in the system whenever phase $\varphi_t$ controlled by the flux piercing the AB ring is detuned from zero. The combination of analytic insights with an unbiased numerics as provided herein opens therefore a path towards optimizing QD-based interferometers for Josephson diode operations.

Finally, we remark that the approach exposed in this work is broadly applicable to multi-terminal devices featuring one or more quantum dots in the central scattering region, connected to multiple superconducting leads. The devices may include multiple loops threaded by magnetic flux.  Our method demonstrates that such systems reduce to simpler effective problems comprising the following elements: the interacting quantum dot(s), the proximitized "side-coupled" levels, and a single continuum of states starting with the lowest superconducting gap among all terminals. The proximitized levels correspond to the Andreev bound states generated by Bogoliubov states being pulled down in the gap region through hybridization. In the presence of the Coulomb repulsion on the quantum dot, if large enough ($U \gtrsim 2\Delta$), Yu-Shiba-Rusinov states will also emerge. This setup thus provides a versatile platform for engineering subgap states with tailored dispersion.

\begin{acknowledgements}
	The authors would like to thank Toma\v{s} Novotn\'{y} for helpful discussions.
    PZ was supported by Grant  No. 23-05263K of the Czech Science Foundation, the project e-INFRA CZ (ID:90140) of the Czech Ministry of Education, Youth and Sports  and the Czech Republic-Germany Mobility programme (ID:8J25DE001).
    DR and RŽ acknowledge the support of the Slovenian Research and Innovation Agency (ARIS) under P1-0416 and J1-3008.
\end{acknowledgements}

\section*{Data availability}
The data that support the findings of this article are openly available \cite{2026-abzenodo}.

\appendix

\begin{widetext}
	
\section{Three-terminal set-up}

The solution for $\mathbb{F}_3$ is
\begin{align}
\mathbb{F}_3
=
\mathbb{G}_{0,d}
\left[
\mathbb{V}_3
+
\mathbb{V}_2\mathbb{G}_2\mathbb{I}_{22}
\left(
\mathbb{T}_{23}
+
\mathbb{T}_{21}\mathbb{G}_1\mathbb{T}_{13}
\right)
+
\mathbb{V}_1
\mathbb{G}_1
\mathbb{T}_{13}
+
\mathbb{V}_1
\mathbb{G}_1
\mathbb{T}_{12} \mathbb{G}_2 \mathbb{I}_{22}
\left(
\mathbb{T}_{23}
+
\mathbb{T}_{21}
\mathbb{G}_1
\mathbb{T}_{13}
\right)
\right]
\mathbb{I}_{33},
\end{align}
where
\begin{align}
\mathbb{I}_{22}
&=
\frac{\mathbb{1}}
{\mathbb{1} - \mathbb{T}_{21}\mathbb{G}_1\mathbb{T}_{12}\mathbb{G}_2},
\\
\mathbb{I}_{33}
&=
\frac{\mathbb{1}}
{
\mathbb{1}
-
\mathbb{T}_{31}\mathbb{G}_1\mathbb{T}_{13}\mathbb{G}_3
-
\left(
\mathbb{T}_{32}+\mathbb{T}_{31}\mathbb{G}_1\mathbb{T}_{12}
\right)
\mathbb{G}_2
\mathbb{I}_{22}
\left(
\mathbb{T}_{23}
+
\mathbb{T}_{21}
\mathbb{G}_{1}
\mathbb{T}_{13}
\right)
\mathbb{G}_3
}.
\end{align}
The remaining quantities $\mathbb{F}_1$ and $\mathbb{F}_2$ are obtained by employing cyclic permutations of $\{1,2,3\}$.


\end{widetext}

\section{Side-coupled mode \label{sec:side_coupled} }

Since the continuous part of the hybridization function has a gap, the newly developed NRG scheme of Ref.~\cite{Zalom-2023} has to be applied. The present case, however, possesses also non-zero pole contributions which have not been discussed in Ref.~\cite{Zalom-2023}. To account for these, we define an electronic mode created (annihilated) by $s^{\dagger}_{\sigma}$ ($s^{\vphantom{\dagger}}_{\sigma}$), where $\sigma \in \{ \uparrow, \downarrow \}$. They form a side-coupled non-interacting dot ($U=0$) with induced pairing $\Delta_s$ and energy level $\varepsilon_s$. Using the Nambu formalism with $S^{\dagger} =  \left(s^{\dagger}_{\uparrow}, s^{\vphantom{\dagger}}_{\downarrow} \right)$, the corresponding Hamiltonian is then defined as
\begin{align}
	H^S
	&=
	S^{\dagger}
	\mathbb{E}_s
	S
	=
	S^{\dagger}
	\left( \Delta_s \sigma_x + \varepsilon_s \sigma_z \right)
	S,
	\label{eq:Hside_nambu_app}
\end{align}
where the induced gap $\Delta_s$ is assumed real and generally different from the underlying BCS gap $\Delta$. The superscript $S$ indicates reference against the spinor basis $S$. The additional mode couples directly only to the interacting QD via
\begin{align}
	H_{T}^S
	&=
	V_{s}
	S^{\dagger}
	\sigma_z
	D
	+
	H.c.
\end{align}
to which contributes the following self-energy to the interacting QD:
\begin{align}
	\mathbb{\Sigma}^{S}
	=
	\frac{V^2_{s}}{z^2-\varepsilon_s^2-\Delta_s^2}
	\left(
	z \mathbb{1}
	-
	\Delta_s \sigma_x
	+
	\varepsilon_s \sigma_z
	\right).
\end{align}
This in turn acquires jumps at $\omega_{s}=\pm\sqrt{\varepsilon_s^2+\Delta_s^2}$, so the corresponding hybridization function becomes
\begin{align}
	\mathbb{D}_{\mathrm{side}}^S(\omega)
	=
	\mathbb{D}^S_{+}(\omega)
	+
	\mathbb{D}^S_{-}(\omega)
\end{align}
with
\begin{align}
	\mathbb{D}^S_{s}(\omega)
	=
	\frac{\pi V^2_{sd}}{2}
	\left(
	\begin{array}{cc}
		1 + \varepsilon_s/\omega_s
		& 
		-\Delta_s/\omega_s
		\\
		-\Delta_s/\omega_s
		& 
		1 - \varepsilon_s/\omega_s
	\end{array}
	\right)
	\delta
	\left(
	{\omega-\omega_{s}}
	\right)
\end{align}
for $s=\pm$. Moreover, note that $\varepsilon_s^2 + \Delta_s^2 =\omega_s^2$ is fulfilled by the SC proximitized side-coupled mode. 

\vspace{5mm}

\begin{figure*}[t]
	\includegraphics[width=2.1\columnwidth]{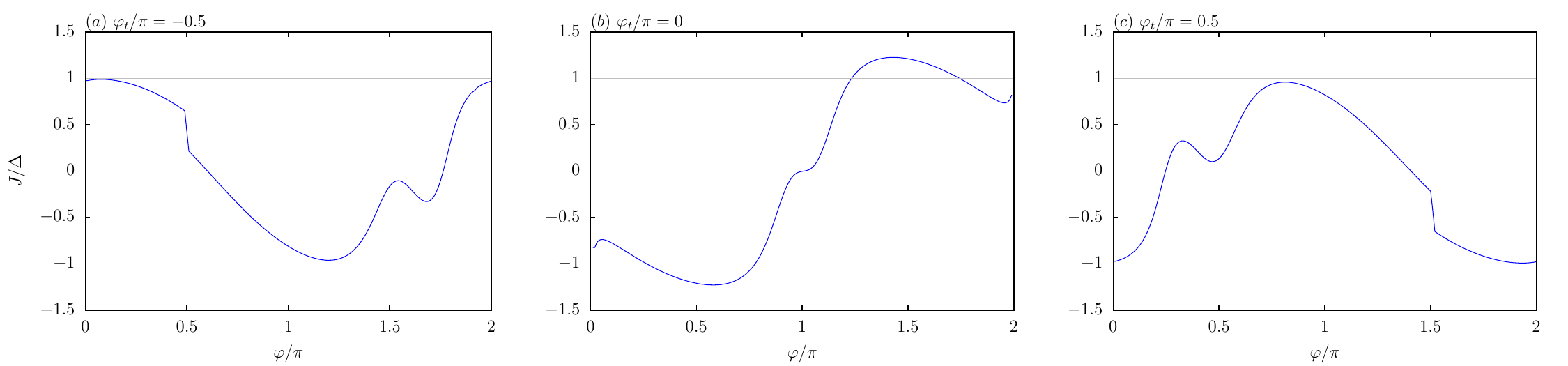}
	\caption{
		Josephson current $J$ calculated in the ABS approximation for $U/\Delta=3$ $\Gamma/\Delta=0.5$ $\tilde{t}=1.5$ at three selected values $\varphi_t/\pi$: in panel (a) $\varphi_t/\pi=-1/2$, in (b)  $\varphi_t/\pi=0$ and in (c) $\varphi_t/\pi=1/2$. Only panels (a) and (c) show the emergence of small differences in forward and backward supercurrents.
These parameters correspond to the spectra presented in Fig.~\ref{fig:abses_phidep}.
		\label{fig:abs_current_phi}}
\end{figure*}

\section{Josephson current \label{sec:josephson} }

In the log-gap NRG scheme the problem is formulated in terms of the Green's function on the quantum dot site. One then constructs a Wilson chain with on-site and hopping coefficients that are explicitly $\varphi$-dependent. Consequently, the Josephson current cannot be obtained by simply taking a derivative with respect to the phase, because the corresponding operator is non-local (it extends over the whole Wilson chain) and its expectation value cannot be easily obtained. Furthermore, the absolute ground state energy depends on the discretization parameters and, in this scheme, this dependence cannot be canceled out by subtracting the absolute ground state of a reference system without the impurity, because the Josephson current is a property of the full system. Instead, Josephson current can be computed by expressing the dot-lead and lead-lead Green's functions in terms of the dot Green's function, then using the approach discussed in Ref.~\cite{Meden-2019}. Here we resort to a calculation that is most reliable in the ABS regime of small $U$, but it is in general only an approximation, where the Josephson current is extracted as $dE_{ex}/d\varphi$ with $E_{ex}$ denoting the energy of the first excited state. The Josephson current in this approximation  is shown in Fig.~\ref{fig:abs_current_phi}. As expected from the $\varphi$-dependence of ABS energies, a small difference between forward and backward current emerges at $\varphi_t/\pi=\pm 0.5$but not at $\varphi_t/\pi=0$. As this is a crude approximation which neglects the continuum contributions, we do not attempt to optimize the model parameters for diode efficiency.


%

\end{document}